\documentclass[twocolumn]{aastex63}

\usepackage{amssymb,amsmath,graphicx,latexsym}
\usepackage{subfigure}

\shorttitle{Primordial black holes formation and secondary gravitational waves ...}
\shortauthors{Teimoori et al.}
\begin{document}

\title{Primordial black holes formation and secondary gravitational waves in nonminimal derivative coupling inflation}

\correspondingauthor{Kayoomars Karami}
\email{kkarami@uok.ac.ir}

% \author[0000-0002-7477-7686]{Zeinab Teimoori}
\author{Zeinab Teimoori}
\affiliation{Department of Physics, University of Kurdistan, Pasdaran Street, P.O. Box 66177-15175, Sanandaj, Iran}

\author{Kazem Rezazadeh}
\affiliation{Department of Physics, University of Kurdistan, Pasdaran Street, P.O. Box 66177-15175, Sanandaj, Iran}
\affiliation{School of Physics, Institute for Research in Fundamental Sciences (IPM),
P.O. Box 19395-5531, Tehran, Iran}

\author{Kayoomars Karami}
\affiliation{Department of Physics, University of Kurdistan, Pasdaran Street, P.O. Box 66177-15175, Sanandaj, Iran}



\begin{abstract}

We study the possibility of the Primordial Black Holes (PBHs) formation with the aim of finding a considerable fraction of Dark Matter (DM), using the gravitationally enhanced friction mechanism which arises from a nonminimal derivative coupling between the scalar field and the gravity. Assuming the nonminimal coupling parameter as a special function of the scalar field and considering the potential of natural inflation, we find three parameter sets that produce a period of ultra slow-roll inflation. This leads to sufficiently large enhancement in the curvature power spectra to form PBHs. We show that under the gravitationally enhanced friction mechanism, PBHs with a mass around ${\cal O}\big(10^{-12}\big)M_\odot$ can constitute around $96\%$ of the total DM and so this class of PBHs can be taken as a great candidate for DM. We further study the secondary Gravitational Waves (GWs) in our setting and show that our model predicts the peak of the present fractional energy density as $\Omega_{GW0} \sim 10^{-8}$ at different frequencies for all the three parameter sets. This value lies well inside the sensitivity region of some GWs detectors at some frequencies, and therefore the observational compatibility of our model can be appraised by the data from these detectors.

\end{abstract}



\section{Introduction} \label{sec:intro}

The idea of Primordial Black Holes (PBHs) formation from the primordial curvature perturbations was first suggested in 1966 by Zel'dovich and Novikov and later developed by Hawking and Carr in the early 1970s \citep{zel1967hypothesis,Hawking:1971ei,Carr:1974nx}. Based on this scenario, the gravitational collapse of overdense inhomogeneities in the very early universe could lead to the formation of PBHs. In recent years, the successful detection of Gravitational Waves (GWs) from merging black holes with mass ${\cal O}(10)M_\odot$ ($M_\odot$ is the solar mass), by LIGO-Virgo Collaboration \citep{Abbott:2016blz,Abbott:2016nmj, Abbott:2017oio,Abbott:2017vtc, abbott2017gw170608} has renewed interest in PBHs and the discussions of PBHs as a Dark Matter (DM) candidate have increased \citep{Bird:2016dcv, Sasaki:2016jop, Blinnikov:2016bxu, Clesse:2016vqa, Clesse:2017bsw}. PBHs as DM at other mass values studied in many works. For example, \citet{mroz2017no, Niikura:2019kqi} have argued that the PBHs with earth mass ($\sim{\cal O}(10^{-5})M_\odot$), can constitute of order ${\cal O}(10^{-2})$ of DM and the ultrashort-timescale microlensing events in the OGLE data can be explained by this class of PBHs. PBHs can constitute all the DM in two asteroid mass intervals: $10^{-16}-10^{-14}M_\odot$ and $10^{-13}-10^{-11}M_\odot$ \citep{Carr:2009jm, Graham:2015apa, Katz:2018zrn, Niikura:2017zjd, Laha:2019ssq, Dasgupta:2019cae}.

Generically, PBHs would form due to the gravitational collapse of the curvature perturbations generated during the inflationary phase, when the relevant scales reenter the horizon in the radiation dominated era. After horizon reentry, if the primordial density perturbations have sufficiently large amplitude, the overdense region would collapse and eventually form PBHs. However, to this aim, there is a big challenge. On one hand, the amplitude of the primordial curvature perturbations at large scales is constrained by the CMB observations to be ${\cal P}_{s}\sim{\cal O}(10^{-9})$ at the pivot scale $k_{*}=0.05 {\rm{Mpc}}^{-1}$ \citep{Akrami:2018odb}. On the other hand, to primordial perturbations lead to form PBHs, it is required that the amplitude of these perturbations be ${\cal P}_{s}\sim{\cal O}(10^{-2})$ at some smaller scales. Therefore, a mechanism is needed to enhance the amplitude of the power spectrum of curvature perturbations by seven orders of magnitude at smaller scales, with respect to the CMB scales. Recently, many works have investigated how such an enhancement could be obtained \citep{Cai:2018tuh, Pi:2017gih, Chen:2019zza, Ballesteros:2018wlw, Kamenshchik:2018sig, Fu:2019ttf, Ashoorioon:2019xqc, Mishra:2019pzq, Dalianis:2019vit}. An appropriate way to this aim is to slow down the scalar field evolution by increasing friction. This leads to an ultra slow-roll inflationary phase which the location and duration of this region depend on the model. In the presence of the nonminimal derivative coupling to gravity and by fine-tuning of the model parameters, this goal can be achieved appropriately, because the nonminimal derivative coupling to gravity provides a gravitationally enhanced friction mechanism \citep{Germani:2010gm, Tsujikawa:2012mk, Tsujikawa:2013ila}.

The nonminimal derivative coupling model is a subset of the Horndeski theory, which is the most general scalar-tensor theory having second-order equations of motion \citep{Horndeski:1974wa,DeFelice:2011uc,Tsujikawa:2012mk, Tsujikawa:2013ila}. This important feature is essential to avoid serious problems such as negative energies and related instabilities \citep{ostrogradski1850malmoires}. An interesting property of the nonminimal derivative coupling scenario is that the mechanism of the gravitationally enhanced friction can be used only for some special potentials \citep{Tsujikawa:2012mk}. Therefore, in this work we consider one of the popular models of inflation, namely natural inflation \citep{Freese:1990rb, Adams:1992bn} which fulfills the basic requirements for the potential of the nonminimal derivative coupling scenario \citep{Tsujikawa:2012mk}. Note that, in the standard inflationary scenario based on the Einstein gravity, the natural potential is in tension with current CMB data, and its results can be compatible with only the 95\% CL of Planck 2018 TT+lowE results \citep{Akrami:2018odb}. This motivates us to investigate whether it is possible to improve its observational results in the setting of nonminimal derivative coupling scenario, such that it be compatible with 68\% CL constraints of these data set. Accordingly, we study the possibility of PBHs formation whose masses are around ${\cal O}(10)M_\odot$, ${\cal O}(10^{-5})M_\odot$, and ${\cal O}(10^{-12})M_\odot$ in the framework of the nonminimal derivative coupling model.

Note that a similar scenario has been considered by \citet{Fu:2019ttf}. In our work, unlike the study of \citet{Fu:2019ttf}, the nonminimal derivative coupling scenario is valid during the whole of inflationary stage. At some scales during inflation, the nonminimal derivative coupling presents a peak which leads to a temporary ultra slow-roll period. This ultra slow-roll period, in turn provides the gravitationally enhanced friction that is required for PBHs production. Away from the ultra slow-roll period, the peak of the nonminimal derivative coupling disappears, but in our model, in contrary to \citet{Fu:2019ttf}, the nonminimal derivative coupling still takes some non-vanishing values. These values result from the existence of another term in the function supposed for the nonminimal derivative coupling which holds all over the inflationary era. As a results, the inflationary observables which are estimated at the epoch of horizon crossing of the curvature perturbations, are affected from the nonminimal derivative coupling, and we hope with the aid of this effect, we can improve the observational results of the natural potential in our setup.

The enhancement of primordial curvature perturbation that leads to the production of PBHs, may also induce the secondary GWs \citep{Matarrese:1997ay, Mollerach:2003nq, Ananda:2006af, Baumann:2007zm,  Saito:2008jc, Saito:2009jt, Bugaev:2009zh, Bugaev:2010bb, Alabidi:2012ex,Nakama:2016gzw, Inomata:2016rbd,Garcia-Bellido:2017aan,Peirone:2017vcq, Cheng:2018yyr, Lin:2020goi}. Therefore, the detection of both PBHs and secondary GWs can be regarded to constrain the substantial increment of the amplitude of the primordial curvature perturbation at some scales during inflation. So, in this way, we can acquire valuable information about the physics of the early universe. So far, various scenarios are proposed to generate secondary GWs from inflation. For instance, in \citet{Saito:2008jc, Bartolo:2018evs, Cai:2018dig}, a phenomenological delta function provides the required enhancement in the scalar power spectrum to produce PBHs and induced GWs. A broken power-law \citep{Lu:2019sti} or Gaussian power spectrum \citep{Namba:2015gja,Garcia-Bellido:2017aan, Lu:2019sti} can also give rise to such an enhancement. Furthermore, other mechanisms such as the running-mass model \citep{Stewart:1996ey,Drees:2011hb,Datta:2019euh}, the axion-curvaton model \citep{Kasuya:2009up, Kawasaki:2012wr}, oscillons after inflation \citep{Easther:2006vd, Antusch:2016con, Liu:2017hua}, and parametric resonance due to the double inflation \citep{Kawasaki:2006zv, Kawasaki:2016pql} can also explain the generation of PBHs and induced GWs in the primordial universe. In this paper, we further investigate the secondary GWs in our inflationary setup. Specifically, we calculate the present fractional energy density of these GWs and compare our findings with the sensitivity regions of different GWs detectors.

The rest of this paper is structured as follows. In Sec. \ref{review:sec}, we propose a brief review of the nonminimal derivative coupling model and find all the necessary equations describing our model. In Sec. \ref{PBH:sec}, applying the Press-Schechter formalism, we review the base formulas for the calculation of the mass and abundance of PBHs first, and then we discuss how to produce the sufficiently large amplitude power spectrum of curvature perturbations with enhanced friction in the framework of the nonminimal derivative coupling scenario. Subsequently, we investigate the secondary GWs in our framework in Sec \ref{sec:sgws}. Finally, we summarize our concluding remarks in Sec. \ref{sec:con}.

\section{Nonminimal Derivative Coupling Model}\label{review:sec}

We start with the nonminimal derivative coupling action \citep{DeFelice:2011uc, Tsujikawa:2012mk, Tsujikawa:2013ila}

\begin{align}
S= & \int d^{4}x\sqrt{-g}\bigg[\frac{1}{2}R-\frac{1}{2}\big(g^{\mu\nu}-g(\phi)G^{\mu\nu}\big)\partial_{\mu}\phi\partial_{\nu}\phi
\nonumber
\\
& -V(\phi)\bigg],
\label{action}
\end{align}
where $g$ is the determinant of the metric $ g_{{\mu}{\nu}}$, $R$ is the Ricci scalar, $G^{{\mu}{\nu}}$ is the Einstein tensor, $g(\phi)$ is the coupling parameter which has a dimension of inverse mass squared, and $V(\phi)$ is the potential of a scalar field $\phi$. Throughout this paper, we take the reduced Planck mass equal to unity, $M_P=(8\pi G)^{-1/2}=1$. As mentioned in Sec. \ref{sec:intro}, the action (\ref{action}) belongs to the Horndeski theory. The Lagrangian in the Horndeski theory contains the terms $K(\phi,X)$, $G_3(\phi,X)\Box \phi$, $G_4(\phi,X)R+G_{4,X}\times$ [field derivative terms], and $G_5(\phi,X)G^{{\mu}{\nu}}(\bigtriangledown_{\mu}\bigtriangledown_{\nu}\phi) -(G_{5,X}/6)\times$  [field derivative terms], where $K$ and $ G_{i} (i=3, 4, 5) $ are general functions of $\phi$ and the kinetic term $X=-\frac{1}{2}g^{\mu\nu}\partial_{\mu}\phi \partial_{\nu}\phi$ and $G_{i,X}=\partial G_{i}/\partial X$ \citep{Tsujikawa:2013ila}. By setting $K(\phi,X)=X-V(\phi)$, $G_3=G_4=0$, and $G_5=G_5(\phi)$, then integrating the term $G_5(\phi)G^{{\mu}{\nu}}\bigtriangledown_{\mu}\bigtriangledown_{\nu}\phi$ by parts and finally defining $g(\phi)\equiv-2 {d G_5}/{d\phi}$, the nonminimal derivative coupling action in Eq.(\ref{action}) is recovered from the Horndeski Lagrangian \citep{Tsujikawa:2013ila}. Note that the case $g=1/M^2$, where $M$ is a constant with a dimension of mass, has been studied in \citet{Germani:2010gm,germani2011uv, DeFelice:2011uc,Tsujikawa:2012mk, Tsujikawa:2013ila} (see also \citet{Amendola:1993uh} for the original work). It is so useful if a more general function is regarded for $g$ such that it depends also on the scalar field $\phi$. In this way, it will be a dynamical quantity that can vary with time during the inflationary period. In this work, we consider this possibility and try to improve the observational results of the natural potential and also provide a reasonable scenario for production of PBHs from inflation, by using a suitable choice for $g(\phi)$.

In the following, let us briefly review the background dynamics and basic formulas governing the theory of cosmological perturbations for the model described by action (\ref{action}). To this aim, we follow the approach of \citet{DeFelice:2011uc,Tsujikawa:2013ila} in which we set $K(\phi,X)=X-V(\phi)$, $G_3=G_4=0$ and $G_5=G_5(\phi)$ to transform their results to be applicable for our model.

From Eqs. (4), (5), (6), and (7) in \citet{DeFelice:2011uc}, one can find the modified
background equations in the flat Friedmann-Robertson-Walker (FRW) metric $g_{{\mu}{\nu}}={\rm diag}\Big(-1, a^{2}(t), a^{2}(t), a^{2}(t)\Big)$, as the following forms

\begin{align}
& 3H^{2}-\frac{1}{2}\Big(1+9H^{2}g(\phi)\Big)\dot{\phi}^{2}-V(\phi)=0,
\label{FR1:eq}
\\
& 2\dot{H}+\left(-g(\phi)\dot{H}+3g(\phi)H^{2}+1\right)\dot{\phi}^{2}-Hg_{,\phi}\dot{\phi}^{3}
\nonumber
\\
& -2Hg(\phi)\dot{\phi}\ddot{\phi}=0,
\label{FR2:eq}
\\
& \left(1+3g(\phi)H^{2}\right)\ddot{\phi}+\Big(1+g(\phi)(2\dot{H}+3H^{2})\Big)3H\dot{\phi}
\nonumber
\\
& +\frac{3}{2}g_{,\phi}H^{2}\dot{\phi}^{2}+V_{,\phi}=0,
\label{Field:eq}
\end{align}
where $H\equiv \dot{a}/a $ is the Hubble parameter and the dot denotes derivative with respect to the cosmic time $t$. We also use the notation $({,\phi})$ to denote ${\partial }/{\partial \phi}$.

In the context of nonminimal derivative coupling scenario, the slow-roll parameters are defined as the following forms \citep{DeFelice:2011uc,Tsujikawa:2013ila}

\begin{align}\label{SRP}
&\varepsilon \equiv -\frac{\dot H}{H^2}, \hspace{.5cm}  \delta_{\phi}\equiv \frac{\ddot{\phi}}{ H\, \dot{\phi}}, \hspace{.5cm}\delta_{X}\equiv \frac{\dot{\phi}^2}{2 H^2}, \hspace{.5cm} \delta_{D}\equiv \frac{g(\phi)\dot{\phi}^2}{4},
 \nonumber
 \\
&\delta_{G}\equiv \frac{g_{,\phi}\dot{\phi}^3}{4H}.
\end{align}
Under the slow-roll approximation, the field equations (\ref{FR1:eq}), (\ref{FR2:eq}), and (\ref{Field:eq}) reduce to

\begin{align}
\label{FR1:SR}
& 3 H^2\simeq V(\phi),\\
  \label{FR2:SR}
& 2\dot{H}+{\cal A}\dot{\phi}^2-H g_{,\phi}\dot{\phi}^3\simeq0,\\
  \label{Field:SR}
& 3 H\dot{\phi}{\cal A}+\frac{3}{2}g_{,\phi}H^2\dot{\phi}^2+V_{,\phi}\simeq0,
\end{align}
where

\begin{equation}\label{A}
{\cal A}\equiv 1+3 g(\phi) H^2.
\end{equation}
For simplicity, we also assume that the following condition holds in the slow-roll regime

\begin{equation}\label{condition}
|g_{,\phi}H\dot{\phi}|\ll {\cal A}.
\end{equation}
Using the condition (\ref{condition}), the field equations (\ref{FR1:SR}), (\ref{FR2:SR}), and (\ref{Field:SR}) can be written in the following forms

\begin{align}
\label{FR1:SRC}
& 3 H^2\simeq V(\phi),\\
  \label{FR2:SRC}
& 2\dot{H}+{\cal A}\dot{\phi}^2\simeq0,\\
  \label{Field:SRC}
& 3 H\dot{\phi}{\cal A}+V_{,\phi}\simeq0.
\end{align}
Taking the time derivative of Eq. (\ref{FR1:SRC}) and using Eq. (\ref{Field:SRC}), we get

\begin{equation}\label{epsilon}
\varepsilon \simeq \delta_{X}+6\delta_{D}\simeq \frac{\varepsilon_{V}}{{\cal A}},
\end{equation}
where

\begin{equation}\label{epsilonv}
\varepsilon_{V}\equiv \frac{1}{2}\left(\frac{V_{,\phi}}{V}\right)^2.
\end{equation}
As we see, $\varepsilon\simeq \varepsilon_{V}$ for ${\cal A}\simeq 1$ and the standard slow-roll inflationary model is recovered. In the high-friction limit $({\cal A}\gg 1)$, one has $\varepsilon \ll \varepsilon_{V}$, which means that the inflaton rolls more slowly down its potential relative to that in standard slow-roll scenario due to a gravitationally enhanced friction.

The power spectrum of the curvature perturbation ${\cal R}$ at the sound exit horizon, i.e. $c_{s}k=a H$ ($k$ is a comoving wavenumber), is given by \citep{ DeFelice:2011uc,Tsujikawa:2013ila}

\begin{equation}\label{Ps}
{\cal P}_{s}=\frac{H^2}{8 \pi ^{2}Q_{s}c_{s}^3}\Big|_{c_{s}k=aH}\,,
\end{equation}
where

\begin{equation}\label{Qs}
 Q_{s}= \frac{ w_1(4 w_{1}{w}_{3}+9{w}_{2}^2)}{3{w}_{2}^2},
\end{equation}

\begin{equation}\label{cs2}
  c_{s}^2=\frac{3(2{w}_{1}^2 {w}_2 H-{w}_{2}^2{w}_4+4{w}_1\dot{w_1}{w}_2-2{w}_{1}^2\dot{w_2})}{{w}_1(4{w}_{1}{w}_{3}+9{ w}_{2}^2)},
\end{equation}
and

\begin{align}
\label{w1}
& {w}_1=1-2\delta_D,\\
  \label{W2}
& {w}_2=2H(1-6\delta_D),\\
& {w}_3=-3 H^2(3-\delta_{X}-36\delta_D),\\
&{w}_4=1+2\delta_D.
\end{align}
Note that, the quantities ${w}_i \; (i=1, 2, 3, 4)$ are obtained by Eqs. (17)-(20) in \citet{Tsujikawa:2013ila} with setting $K(\phi,X)=X-V(\phi)$, $G_3=G_4=0$, and $G_5=G_5(\phi)$. The observational value of the amplitude of scalar perturbations at the CMB pivot scale $k_{*}=0.05\,{\rm Mpc}^{{\rm -1}}$ measured by the Planck team is ${\cal P}_{s}(k_{*})\simeq 2.1 \times 10^{-9}$ \citep{Akrami:2018odb}.

With the help of Eqs. (6), (22), (23), (24), and (25) in \citet{Tsujikawa:2013ila}, and also using Eq. (\ref{epsilon}) to the lowest order in the slow-roll parameters, one can find

\begin{equation}\label{Qs:SR}
 Q_{s}\simeq \delta_{X}+6\delta_{D}\simeq \varepsilon \simeq \frac{\varepsilon_V}{{\cal A}},
\end{equation}
and

\begin{equation}\label{cs2:SR}
  c_{s}^2\simeq 1-\frac{2\delta_D(3\delta_{X}+34\delta_D-2\delta_{\phi})}{\delta_{X}+6\delta_D}.
\end{equation}
Since $ c_{s}^2=1-{\cal O}(\varepsilon)$, applying Eqs. (\ref{FR1:SRC}), (\ref{epsilonv}), and the last approximate equality in Eq. (\ref{Qs:SR}), the power spectrum (\ref{Ps}) at the leading order in the slow-roll parameters, can be obtained as

\begin{equation}\label{PsSR}
{\cal P}_{s}\simeq \frac{V^3}{12\pi^2 V_{,\phi}^2}{\cal A}\simeq\frac{V^3}{12\pi^2 V_{,\phi}^2}\Big(1+g(\phi)V\Big)\,,
\end{equation}
where in the last approximate equality, we have used the definition ${\cal A}$ in Eq. (\ref{A}) and the first Friedmann equation (\ref{FR1:SRC}).

During slow-roll inflationary phase, the Hubble parameter $H$ and the sound speed $c_s$ change much slower than the scale factor $a$ of the universe \citep{garriga1999perturbations}. Therefore, applying the relation $c_s k=a H$ which is valid around the sound horizon exit, we can write $d\ln k\approx H dt$. Using this approximation and the definition $n_s-1 \equiv d\ln{\cal P}_{s}/d\ln k$, one can easily find $n_s-1\simeq \dot{{\cal P}_{s}}/(H{\cal P}_{s})$. From this relation and then using Eqs. (\ref{FR1:SRC}), (\ref{Field:SRC}), (\ref{epsilonv}), and the first approximate equality in (\ref{PsSR}), we can find the scalar spectral index $n_s$ as

\begin{align}\label{nsSR}
n_s-1\simeq & -\frac{1}{{\cal A}}\left[6\varepsilon_{V}-2\eta_{V}+2\varepsilon_{V}\left(1-\frac{1}{{\cal A}}\right)\right.
\nonumber
\\
& \left.
\times \left(1+\frac{g_{,\phi}}{g(\phi)}\frac{V(\phi)}{V_{,\phi}}\right)\right],
\end{align}
where

\begin{equation}\label{eta}
\eta_{V}=\frac{V_{,\phi\phi}}{V}.
\end{equation}
The observational constraint on the scalar spectral index is $n_{s}=0.9627\pm0.0060$ (68\% CL, Planck 2018 TT+lowE) \citep{Akrami:2018odb}.

Using the approximation $d\ln k\approx H dt$ and Eq. (\ref{Field:SRC}), one can obtain the running of the scalar spectral
index as
\begin{equation}\label{alphas}
\frac{d{n_s}}{d\ln k}\simeq -\frac{1}{{\cal A}}\left(\frac{V_{,\phi}}{V(\phi)}\right) n_{s,\phi}.
\end{equation}
The running of the spectral index measured by the Planck team is about $d{n_s}/d \ln k= - {\rm{0}}{\rm{.0078}} \pm 0.0082$ (68\% CL, Planck 2018 TT+lowE) \citep{Akrami:2018odb}.

The tensor power spectrum in the framework of nonminimal derivative coupling scenario is given by \citep{DeFelice:2011uc,Tsujikawa:2013ila}

\begin{equation}\label{Pt}
{\cal P}_{t}=\frac{H^2}{2\pi ^{2}Q_{t}c_{t}^{3}}\Big|_{c_{t}k=aH}\,,
\end{equation}
where

\begin{equation}\label{Qt}
Q_t=\frac{1}{4}{w}_1=\frac{1}{4}(1-2\delta_D),
\end{equation}

\begin{equation}\label{ct2}
c_{t}^2=\frac{{w}_4}{{w}_1}=1+4\delta_{D}+{\cal O}(\varepsilon^2).
\end{equation}
At leading order in slow-roll parameters, the tensor power spectrum (\ref{Pt}) can be expressed as

\begin{equation}\label{PtSR}
{\cal P}_{t}=\frac{2H^2}{\pi ^{2}}\,,
\end{equation}
which is evaluated at ${c_{t}k=aH}$. From Eqs. (24) and (34) in \citet{Tsujikawa:2013ila}, and also using Eqs. (\ref{epsilon}) and (\ref{cs2:SR}), the tensor-to-scalar ratio in the slow-roll regime is obtained as

\begin{equation}\label{r}
r\simeq 16 \varepsilon \simeq 16 \frac{\varepsilon_V}{{\cal A}}.
\end{equation}
The Planck 2018 data provides an upper bound on the tensor-to-scalar ratio as $< 0.0654
$ (68\% CL, Planck 2018 TT+lowE) \citep{Akrami:2018odb}. It is worth mentioning that for the case $g(\phi)= \textrm{const.}$, Eqs. (\ref{nsSR}) and (\ref{r}) are transformed to the corresponding results obtained in \citet{Tsujikawa:2012mk}.

\section{Primordial Black Holes formation}\label{PBH:sec}

In this section, at first, we briefly study the basic formalism for the PBHs formation, and then we apply this formalism in the framework of the nonminimal derivative coupling scenario to investigate the possibility of the PBHs formation.

\subsection{Basic Formalism}

The scalar modes of perturbations reenter the horizon during the radiation dominated era, after inflation. If the amplitude of these perturbations is large enough, they may collapse to form PBHs. The mass of formed PBHs is proportional to the horizon mass as \citep{Gong:2017qlj}

\begin{align}
M(k) & =\gamma M_{hor}
\nonumber
\\
& =3.68\left(\frac{\gamma}{0.2}\right)\left(\frac{g_{*}}{10.75}\right)^{-1/6}\left(\frac{k}{10^{6}\,{\rm Mpc^{-1}}}\right)^{-2}M_{\odot},
\label{mass}
\end{align}
where $\gamma$ is the efficiency of collapse and assumed to be $\gamma \simeq 0.2 $ \citep{carr1975primordial}, $M_{hor}$ is the horizon mass, and $g_*=107.5$ indicates the effective degrees of freedom for energy density at the formation of PBHs.

Assuming that the curvature perturbations obey Gaussian statistics, the fractional energy density of PBHs with mass $M(k)$ at formation can be calculated with the help of Press-Schechter formalism as \citep{Press:1973iz, Young:2014ana, ozsoy2018mechanisms, Tada:2019amh}

\begin{equation}\label{beta}
\beta(M)=\sqrt{\frac{2}{\pi}} \frac{\sigma(M)}{\delta_c}\exp{\left(-\frac{\delta_c^{2}}{2\sigma^2(M)}\right)},
\end{equation}
where $\delta_c$ denotes the threshold of the density perturbation for the PBHs formation. As suggested in \citet{Musco:2012au, Harada:2013epa, Escriva:2019phb, Lin:2020goi}, we consider the threshold value $\delta_c\simeq0.4$. $\sigma^2\big(M(k)\big)$ that is the coarse-grained variance of the density contrast smoothed on a scale $k$ is given by \citep{Young:2014ana, ozsoy2018mechanisms}

\begin{equation}\label{sigma2}
\sigma^2(k)=\int{\frac{dq}{q}}\,W^2(q/k)\frac{16}{81}(q/k)^4 {\cal P}_s(q),
\end{equation}
where ${\cal P}_s(k)$ is the power spectrum of the curvature perturbations and $W(x)$ is the window function. In this work, we use the smoothing Gaussian window as $W(x)=\exp{(-x^2/2)}$.

The current energy fraction of PBHs $(\Omega_{\rm {PBH}})$ to dark matter ($\Omega_{\rm{DM}})$ is defined as \citep{Carr:2016drx,Gong:2017qlj}

\begin{align}\label{fPBH}
f_{\rm{PBH}}(M)\equiv \frac{\Omega_{\rm {PBH}}}{\Omega_{\rm{DM}}}= & \frac{\beta(M)}{3.94\times10^{-9}}\left(\frac{\gamma}{0.2}\right)^{1/2}\left(\frac{g_*}{10.75}\right)^{-1/4}
\nonumber
\\
& \times\left(\frac{0.12}{\Omega_{\rm{DM}}h^2}\right)
\left(\frac{M}{M_{\odot}}\right)^{-1/2}.
\end{align}
Here, we take $\Omega_{\rm {DM}}h^2\simeq0.12$ from the Planck 2018 results \citep{Akrami:2018odb}.

As we mentioned in Sec. \ref{sec:intro}, in this work we consider three distinct PBH mass scales: ${\cal O}(10)M_\odot$, ${\cal O}(10^{-5})M_\odot$, and ${\cal O}(10^{-12})M_\odot$. With the help of Eq. (\ref{mass}), one can easily find that these masses are corresponding to the comoving wavenumber ${\cal O}(10^5)$, ${\cal O}(10^{8})$, and ${\cal O}(10^{12})$ $\rm{Mpc}^{-1}$, respectively.

\subsection{Primordial Black Holes from Nonminimal Derivative Coupling Model}

Now we are in a position to study the possibility of PBHs formation in our model. From Eq. (\ref{PsSR}), it is realized that the power spectrum of the curvature perturbations gets amplified if the coupling parameter $g(\phi)$ has a sufficiently large peak about a field value. Therefore, to obtain a desirable power spectrum with a peak on a particular scale which can lead to the formation of PBHs, we take the functional form of $g(\phi)$ as

\begin{equation}\label{g}
g(\phi)=g_I(\phi)\Big(1+g_{II}(\phi)\Big),
\end{equation}
where

\begin{equation}\label{gI}
g_I(\phi)=\frac{1}{M^{\alpha+1}}\phi^{\alpha-1},
\end{equation}
and

\begin{equation}\label{gII}
g_{II}(\phi)=\frac{\omega}{\sqrt{\left(\frac{\phi-\phi_c}{\sigma}\right)^2+1}}\,.
\end{equation}
In our work, the coupling function $g(\phi)$ consist of the two functions $g_I(\phi)$ and $g_{II}(\phi)$. The function $g_{II}(\phi)$ have a peak at the critical field value $\phi = \phi_c$, and the height and width of the peak are specified by the parameters $\omega$ and $\sigma$, respectively. For the field values away from $\phi = \phi_c$, the function $g_{II}(\phi)$ almost vanishes, so that $g(\phi)\approx g_I(\phi)$. The function $g_I(\phi)$ is a conceivable generalization for the original coupling $1/M^2$ which was first proposed by \citet{Germani:2010gm}, and studied later in the literature \citep{DeFelice:2011uc,Tsujikawa:2012mk, Tsujikawa:2013ila}. More precisely, one can recover this term by setting $\alpha = 1$ in the expression \eqref{gI} for $g_I(\phi)$. In the paper by \citet{Fu:2019ttf}, the coupling function $g(\phi)$ consist of only the one function with the form given in Eq. \eqref{gII}. Therefore, away from the peak, the coupling function $g(\phi)$ disappears and the nonminimal derivative scenario is no longer valid, and we recover the Einstein general relativity. But, in our work, since $g_I(\phi)$ has a non-vanishing value in the regions away the peak, therefore the scenario of nonminimal derivative coupling remains valid in all the regions of the field excursion. Actually, the validity of this scenario in those regions helps us to improve the consistency of the natural potential results for the inflationary observables with the latest CMB data \citep{Akrami:2018odb}. In other words, in our work the scenario of nonminimal derivative coupling is valid throughout the inflationary era, while in \citet{Fu:2019ttf}, this scenario is only valid around $\phi = \phi_c$ where the coupling function has a considerable value, and in other regions the Einstein gravity is recovered, because the nonminimal coupling term vanishes.

We consider the natural potential which has the following form \citep{Freese:1990rb, Adams:1992bn}

\begin{equation}\label{potential}
V(\phi)=\Lambda^4\left(1+\cos\big(\frac{\phi}{f}\big)\right),
\end{equation}
where $\Lambda$ and $f$ are constants with dimensions of mass. In the framework of standard inflationary scenario, the consistency of the natural potential with the latest observations is not very good, because its results can satisfy only the 95\% CL constraints of Planck 2018 TT+lowE data \citep{Akrami:2018odb}. This motivates us to investigate natural inflation in the nonminimal derivative coupling setting to see whether its predictions can be improved in light of the observational results. It seems that, the regime $f\gtrsim M_P$ is impossible in string theory \citep{Banks:2003sx, Barnaby:2010vf}. Since a global symmetry related to pseudo-Nambu-Goldstone Boson (pNGB) is broken above the quantum gravity scale and consequently, the effective field theory may not be valid.

Using Eqs.(\ref{g}), (\ref{gI}), (\ref{gII}), and (\ref{potential}), the power spectrum (\ref{PsSR}) is simplified to

\begin{align}
{\cal P}_{s}\simeq & \frac{f^{2}\Lambda^{4}}{12\pi^{2}}\,\csc^{2}\chi\big(1+\cos\chi\big)^{3}
\nonumber
\\
& \times\Bigg[1+\frac{\lambda\chi^{-1+\alpha}}{f^{2}}\Big(1+\frac{\omega}{\sqrt{1+(\frac{f(\chi-\chi_{c})}{\sigma})^{2}}}\Big)\Bigg],
\label{PsNatural}
\end{align}
where $\chi=\phi/f$, $\chi_c=\phi_c/f$ and

\begin{equation}\label{lambda}
\lambda\equiv\frac{f^{1+\alpha} \Lambda^4}{M^{1+\alpha}}.
\end{equation}
Now, we assume that the following conditions are satisfied:

\begin{equation}\label{assumptions}
\omega\gg 1, \hspace{1.5cm} \mid \phi_{*}-\phi_{c}\mid\, \gg \sigma \omega , \hspace{1.5cm} \lambda \gg f^2,
\end{equation}
where $\phi_{*}$ is the scalar field value at the moment of horizon crossing. Using Eq. (\ref{PsNatural}) and the three assumptions in (\ref{assumptions}), one can easily find

\begin{align}\label{Pspeak}
{\cal P}_{s}\Big|_{\chi= \chi_{c}}\simeq & \; \omega\left(\frac{1+\cos\chi_{c}}{1+\cos\chi_{*}}\right)^{3}\left(\frac{\csc\chi_{c}}{\csc\chi_{*}}\right)^{2}\left(\frac{\chi_{c}}{\chi_{*}}\right)^{-1+\alpha}
\nonumber
\\
& \times{\cal P}_{s}\Big|_{\chi=\chi_{*}},
\end{align}
which the amplitude of the power spectrum at the pivot scale $k_{*}=0.05 {\rm {Mpc}}^{-1}$ is constrained by the Planck 2018 results as ${\cal P}_s\Big|_{\chi=\chi_*}\simeq 2.1 \times 10^{-9}$ \citep{Akrami:2018odb}.

The anisotropies observed in the CMB radiation exit the Hubble horizon around $N_\ast \approx 50-60$ before the end of inflation \citep{Liddle:2003as, Dodelson:2003vq}. The exact value of horizon exit $e$-fold number depends on energy scale of inflation and reheating mechanism after inflation \citep{Liddle:2003as, Dodelson:2003vq}. In our model, like most of conventional inflationary models, inflation occurs at the GUT energy scale ($10^{16}\,\mathrm{GeV}$). Also, in our model, again like most of conventional inflationary models, the details of reheating process after inflation is still unknown for us, and therefore, we cannot determine definitively the exact value of $N_\ast$ in our model. However, we here take $N_\ast = 60$ since it provides better consistency for the inflationary observables in our investigation. Although, we can get almost similar results for the PBHs parameters with the choice of other values of $N_\ast$ in the range $50-60$, but it hurts the consistency of the inflationary observables with the CMB observations in our model.

From Eq. (\ref{Pspeak}) it is clear that the power spectrum at $\chi=\chi_c$ will have a peak of order ${\cal O}(10^{-2})$, provided that $\omega \sim {\cal O}(10^7)$. With this in mind and based on the conditions (\ref{assumptions}), we chose $\alpha=0$, $f=0.2$, and $\lambda=20$. Moreover, we consider three distinct parameter sets which are summarized in Table \ref{Table1}.

\begin{table*}[ht!]
  \centering
  \caption{The chosen parameter sets corresponding to Case 1, Case 2, and Case 3. The value of $\Lambda$ is fixed by imposing the CMB normalization at $N_*=60$.}
\scalebox{1}[1] {
    \begin{tabular}{ccccc}
    \hline
    \hline
    $\qquad \# \qquad$ & $\qquad \phi_{c} \qquad$ & $\qquad \qquad \omega \qquad \qquad$ & $\qquad \qquad \sigma \qquad \qquad$ & $\qquad \Lambda \qquad$\tabularnewline
    \hline
    Case 1 & $0.233760$ & $6.22820\times10^{7}$ & $5.5930\times10^{-11}$ & $0.0050$\tabularnewline
    \hline
    Case 2 & $0.198752$ & $5.00780\times10^{7}$ & $4.2080\times10^{-11}$ & $0.0049$\tabularnewline
    \hline
    Case 3 & $0.176323$ & $4.00821\times10^{7}$ & $3.8180\times10^{-11}$ & $0.0048$\tabularnewline
    \hline
    \end{tabular}
    }
  \label{Table1}
\end{table*}

Note that our model is characterized by eight free parameters $\{\alpha, M, f, \lambda, \Lambda, \omega, \phi_c, \sigma\}$ which the parameters $\alpha$, $M$, $f$, $\lambda$, and $\Lambda$ are related to each other according to Eq. (\ref{lambda}). The value of the other parameter $\Lambda$ is fixed by imposing the CMB normalization at the observable scale. The derived values of the inflationary observables $n_s$, $r$, $d{n_s}/d\ln k$, and also the quantities relevant for producing PBHs of these three cases are presented in Table \ref{Table2}.

\begin{table*}[ht!]
  \centering
  \caption{Results obtained for the three cases of Table \ref{Table1}. We give the scalar spectral index ($n_s$), the tensor-to-scalar ratio ($r$), the running of the scalar spectral index ($dn_s/d\ln k$) and the PBHs fractional abundance ($f_{\rm{PBH}}^{\rm{peak}}$). $M_{\rm{PBH}}^{\rm{peak}}$ is the PBH mass corresponding to $f_{\rm{PBH}}^{\rm{peak}}$.
  The inflationary observables $n_s$, $r$, and $dn_s/d\ln k$ are computed at horizon crossing $e$-fold number $N_{*}=60$. }
\scalebox{1}[1] {\begin{tabular}{c c c c c c }
    \hline
    \hline
     \#  & \,\,\,\,\,\,\,\,\,$n_s$ &\,\,\,\,\,\,\,\,\,  $r$  &\,\,\,\,\,\,\,\,\, $dn_s/d\ln k$ &\,\,\,\,\,\,\,\,\,  $M_{\rm{PBH}}^{\rm{peak}}/M_\odot$ &\,\,\,\,\,\,\,\,\, $f_{\rm{PBH}}^{\rm{peak}}$\\
    \hline
    Case 1 &\,\,\,\,\,\,\,\,\, $0.9527$ &\,\,\,\,\,\,\,\,\, $0.0342$& \,\,\,\,\,\,\,\,\,  $-0.0010$  &\,\,\,\,\,\,\,\,\,$1.47\times10^{-12}$ &\,\,\,\,\,\,\,\,\,$0.9563$\\
    Case 2 &\,\,\,\,\,\,\,\,\, $0.9560$ &\,\,\,\,\,\,\,\,\, $0.0320$&\,\,\,\,\,\,\,\,\,  $ -0.0006$ &\,\,\,\,\,\,\,\,\,$1.64\times10^{-5}$ &\,\,\,\,\,\,\,\,\, $0.0393$\\
    Case 3 &\,\,\,\,\,\,\,\,\, $0.9645$ &\,\,\,\,\,\,\,\,\, $0.0296$& \,\,\,\,\,\,\,\,\,  $0.0013$  &\,\,\,\,\,\,\,\,\,$25.70$ &\,\,\,\,\,\,\,\,\,$0.0018$\\
    \hline
    \end{tabular}}
  \label{Table2}
\end{table*}

\begin{figure*}
\begin{minipage}[b]{1\textwidth}
\subfigure{\includegraphics[width=.48\textwidth]%
{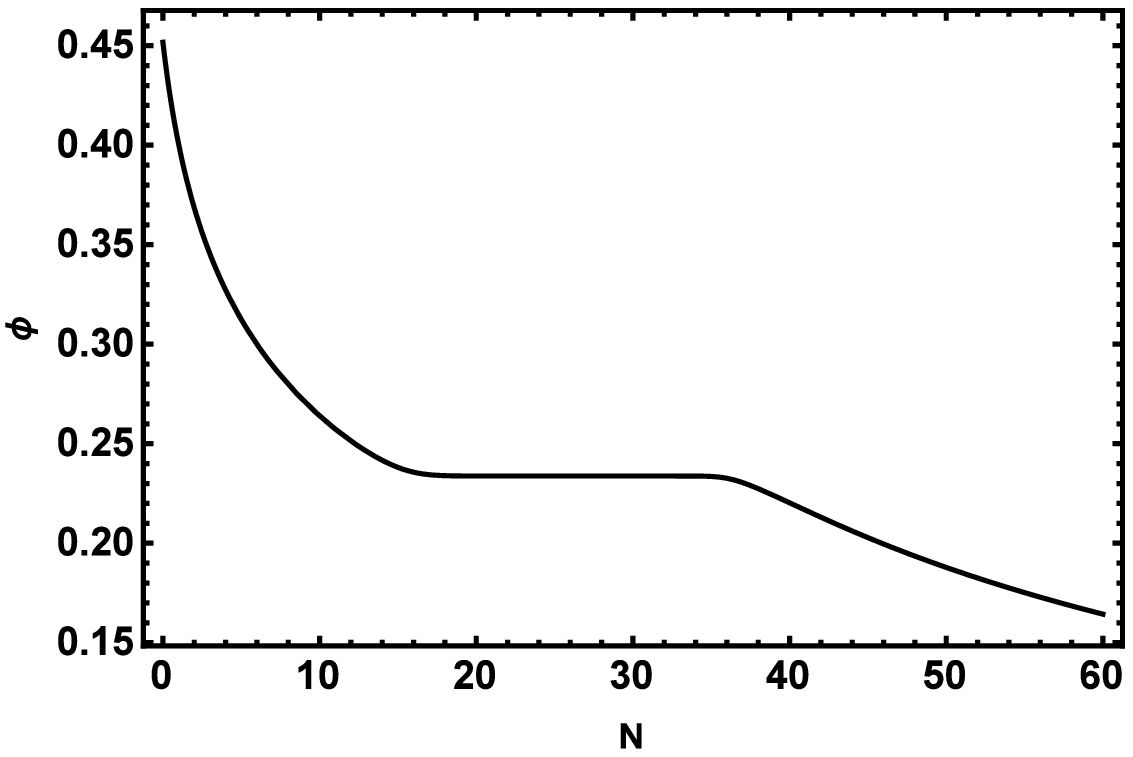}} \hspace{.1cm}
\subfigure{ \includegraphics[width=.48\textwidth]%
{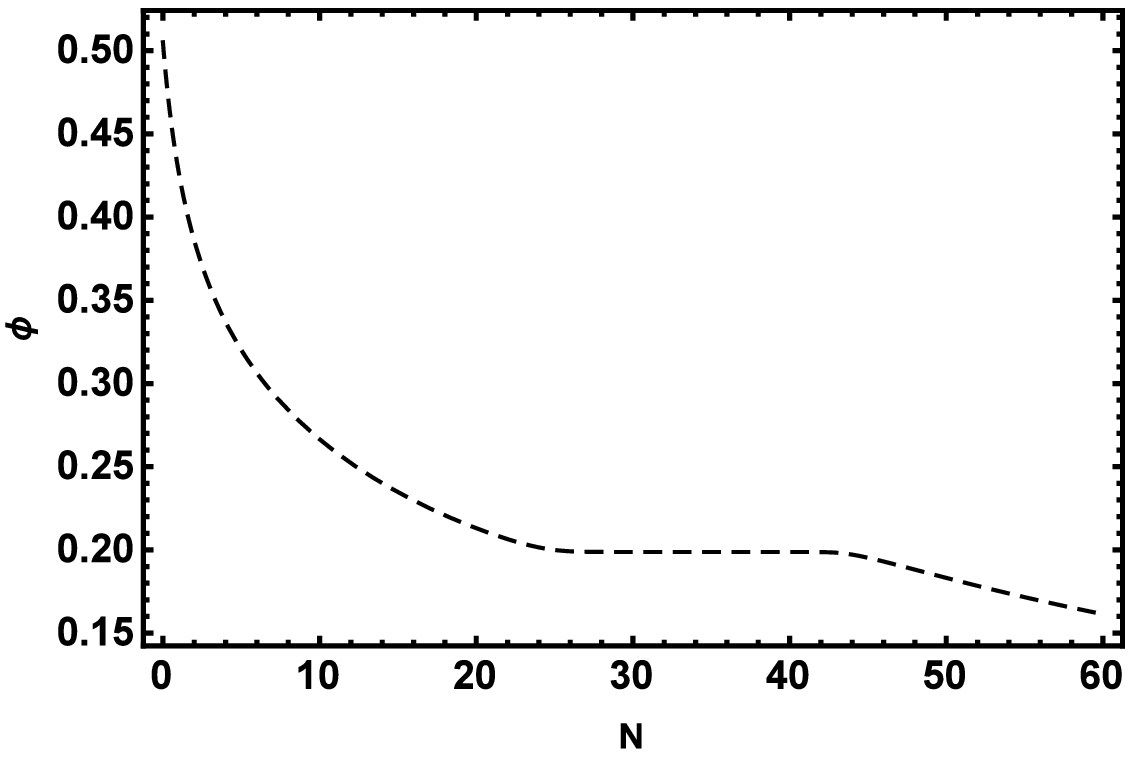}}\hspace{.1cm}
\subfigure{ \includegraphics[width=.48\textwidth]%
{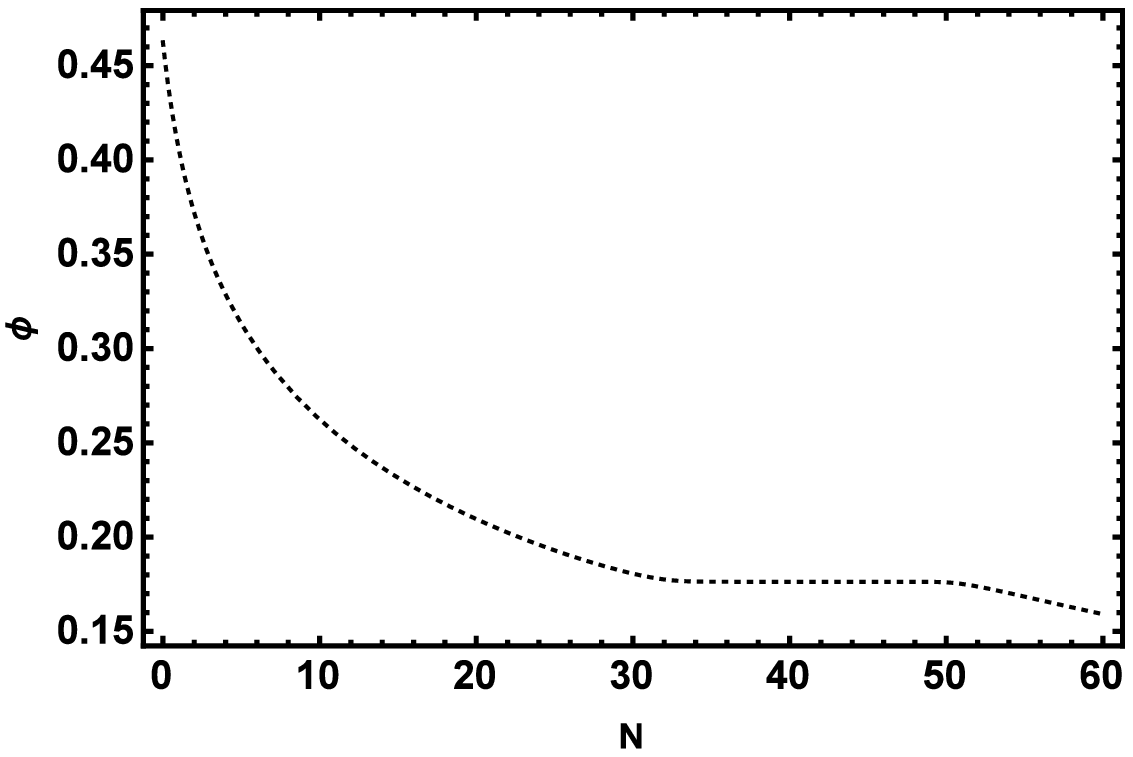}}
\end{minipage}
\caption{Evolution of the scalar field $\phi$ as a function of the $e$-fold number $N$ for Case 1 (solid line), Case 2 (dashed line), and Case 3 (dotted line). The initial conditions are found by using Eqs. (\ref{FR1:SRC}) and (\ref{Field:SRC}) at $N_{*}=60$.}
\label{fig:phi}
\end{figure*}

\begin{figure*}[t]
\begin{center}
\scalebox{0.9}[0.9]{\includegraphics{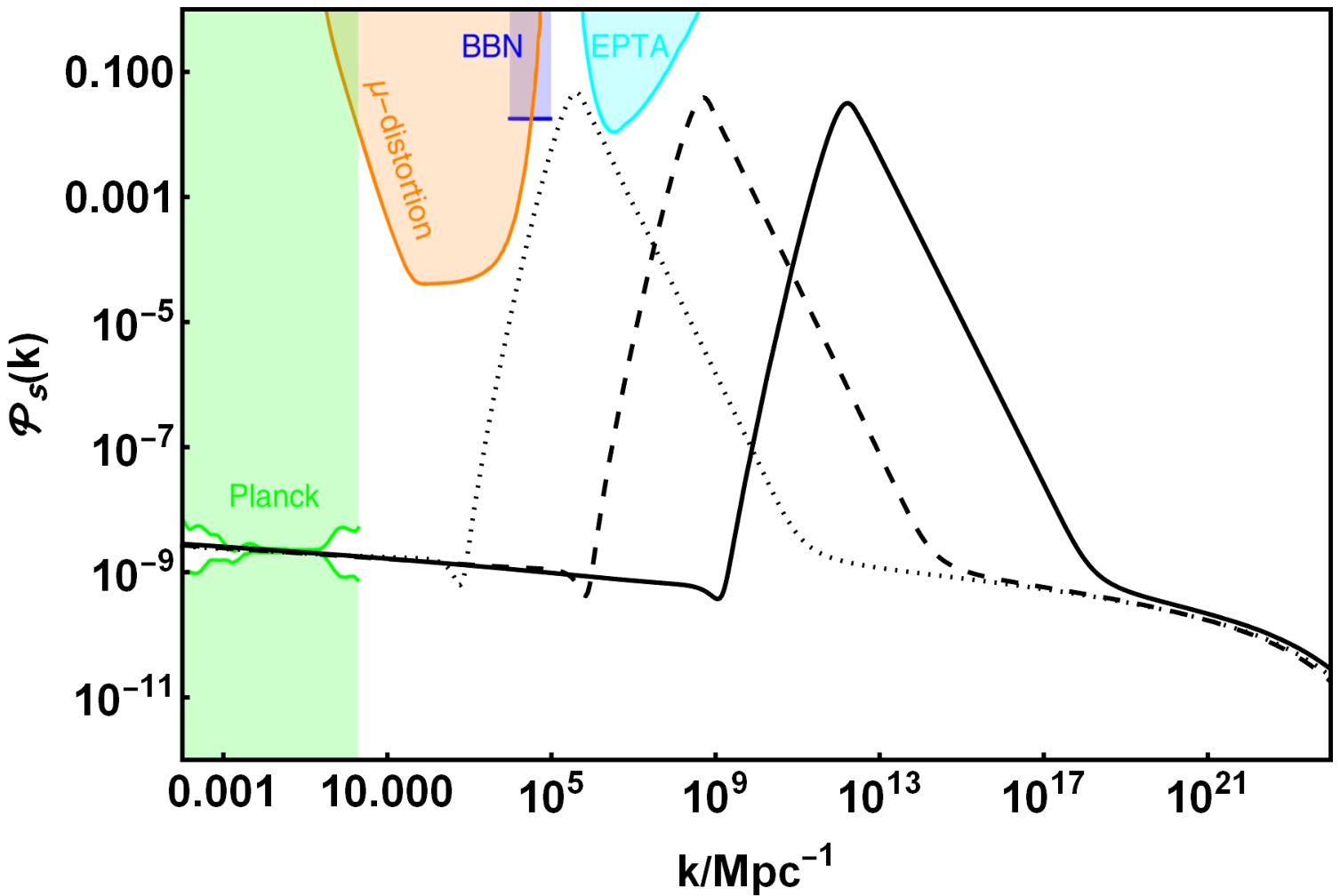}}
\caption{The curvature power spectra found by solving the Mukhanov-Sasaki equation, numerically, as a function of the comoving wavenumber $k$ for Case 1 (solid line), Case 2 (dashed line), and Case 3 (dotted line). The light green shaded region shows the CMB observations \citep{Akrami:2018odb}. The orange, blue, and cyan shaded regions indicate the constraints on the power spectrum from the $\mu$ distortion of CMB \citep{Fixsen:1996nj}, the effect on the ratio between neutron and proton during the big bang nucleosynthesis (BBN) \citep{Inomata:2016uip}, and the current PTA observations \citep{Inomata:2018epa}, respectively.}
\label{fig:Psk}
\end{center}
\end{figure*}

Solving the full equations of motion (\ref{FR1:eq})-(\ref{Field:eq}), for the potential (\ref{potential}) and the nonminimal derivative coupling parameter given in Eqs. (\ref{g})-(\ref{gII}) numerically, we plot the evolution of the scalar field $\phi$ as a function of $e$-fold number $N$ (where $dN=-Hdt$) in Figure  \ref{fig:phi} for Case 1 (solid line), Case 2 (dashed line), and Case 3 (dotted line). In this figure, we see that a plateau-like region appears in the $e$-folds $17 \lesssim N\lesssim 36$ for Case 1, in $26 \lesssim N\lesssim 44$ for Case 2, and in $33\lesssim N\lesssim 50$ for Case 3. This behavior in the evolution of $\phi$ versus $N$ arises from the slowing down of the speed of the scalar field around $\phi = \phi_c$. The presence of this region leads to a large enhancement of the curvature power spectrum (see Figure  \ref{fig:Psk}). The plateau-like region exhibiting a phase of ultra slow-roll inflation in which inflaton rolls very slowly due to the high friction.

\begin{figure*}[t]
\begin{center}
\scalebox{0.45}[0.45]{\includegraphics{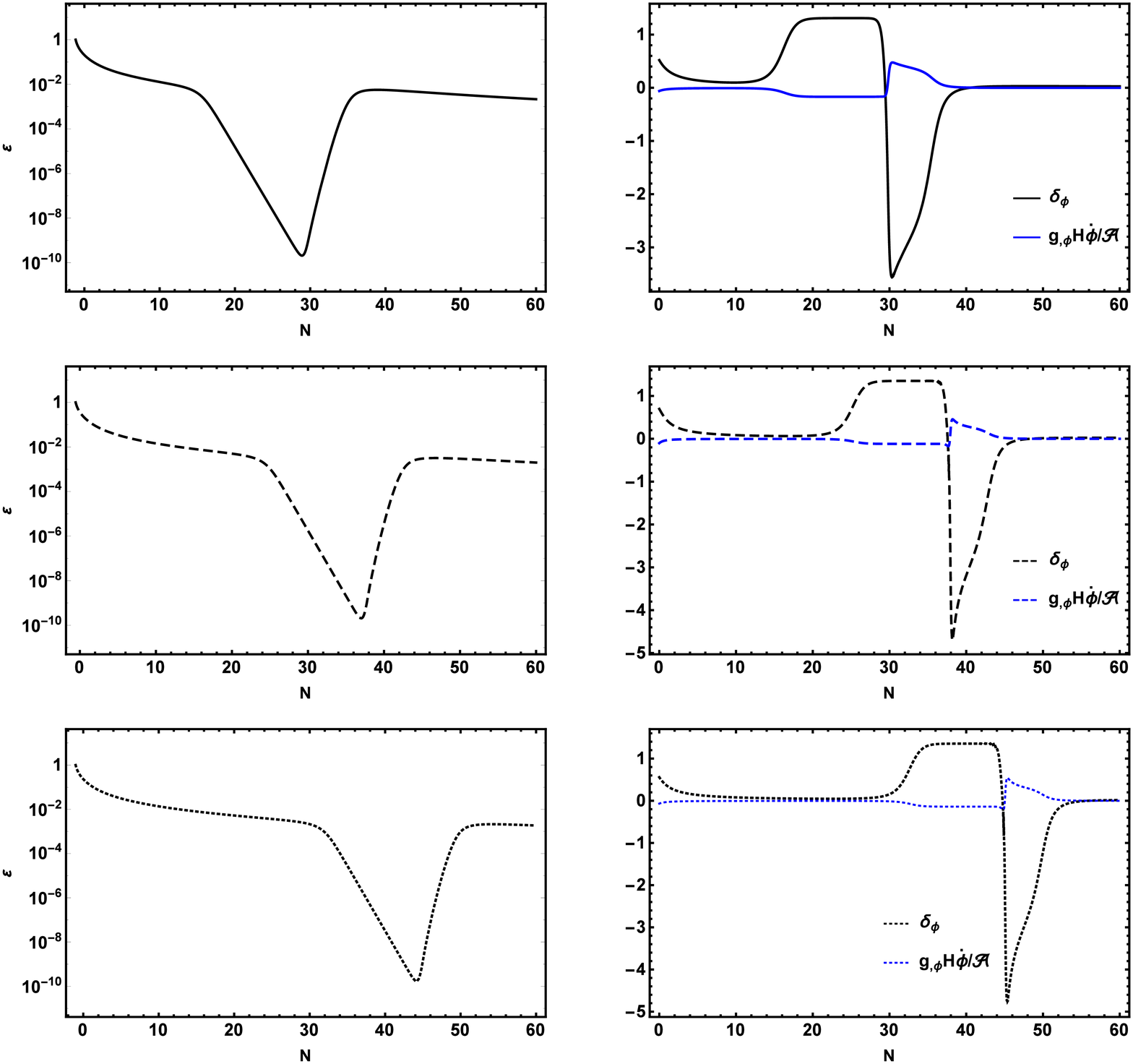}}
\caption{Evolution of the first slow-roll parameter $\varepsilon$ (left), the second slow-roll parameter $\delta_{\phi}$, and the expression ${g_{,\phi}H\dot{\phi}}/{\cal A}$ (right) versus the $e$-fold number $N$ for Case 1 (solid line), Case 2 (dashed line), and Case 3 (dotted line).}
\label{fig:SRparameters}
\end{center}
\end{figure*}

In Figure  \ref{fig:SRparameters}, the evolution of the first slow-roll parameter $\varepsilon$ (left panel), the second slow-roll parameter $\delta_\phi$, and the expression ${g_{,\phi}H\dot{\phi}}/{\cal A}$ (right panel) are plotted as a function of $N$ for the three cases of Table \ref{Table1}. The left panel of this figure clears that $\varepsilon$ remains below unity up until the end of inflation. Of course, it does not become unity at $N=0$ and inflation continues for a few more $e$-folds (1.159 $e$-folds for Case 1, 0.557 $e$-folds for Case 2, and 1.008 $e$-folds for Case 3). The reason of this behavior is the strong slowing down of inflaton during the ultra slow-roll phase. However, since in our model inflation ends in a few $e$-folds after $N=0$, it ensures us that inflation ends after a finite duration.
From the right panel of this figure, we further see that the slow-roll condition $\mid \delta_{\phi}\mid\ll1$ is violated at the ultra slow-roll period.

From Figure  \ref{fig:SRparameters}, one can easily find that at the time of the sound horizon corresponding to $N_{*}=60$, the slow-roll approximation remains valid. It is worth mentioning that under the slow-roll limit, both two parameters $\varepsilon$ and $\delta_{\phi}$ should be less than unity. The condition (\ref{condition}) is also valid at $N_{*}=60$. Therefore, by using Eqs. (\ref{A}), (\ref{epsilonv}), (\ref{nsSR}), (\ref{eta}), (\ref{alphas}), and (\ref{r}), we can find the scalar spectral index $n_s$, the tensor-to-scalar ratio $r$, and the running of the scalar spectral index $dn_s/d\ln k$  for the model described by Eqs. (\ref{g})-(\ref{potential}).
The numerical results are presented in Table \ref{Table2}, indicate that the values of $r$, and $dn_s/d\ln k$ predicted by our model in Case 1 and Case 2, are consistent with the $68\%$ CL constraints of Planck 2018 TT+lowE data \citep{Akrami:2018odb}, and
the values of $n_s$ satisfy the $95\%$ CL constraints.
For Case 3, the value of $dn_s/d\ln k$ is in agreement with the 95\% CL region of Planck 2018 TT+lowE data \citep{Akrami:2018odb}, while the values of $n_s$ and $r$ satisfy the $68\%$ CL constraints of these data. Therefore, we see that in the framework of the nonminimal derivative coupling, we can improve the consistency of the inflationary observables of the natural potential in light of the Planck 2018 results \citep{Akrami:2018odb} in comparison with its results in the setup of standard inflationary scenario. This is a remarkable consequence of the consideration of the nonminimal derivative coupling scenario in our investigation.

\begin{figure*}
\begin{minipage}[b]{1\textwidth}
\subfigure{\includegraphics[width=.48\textwidth]%
{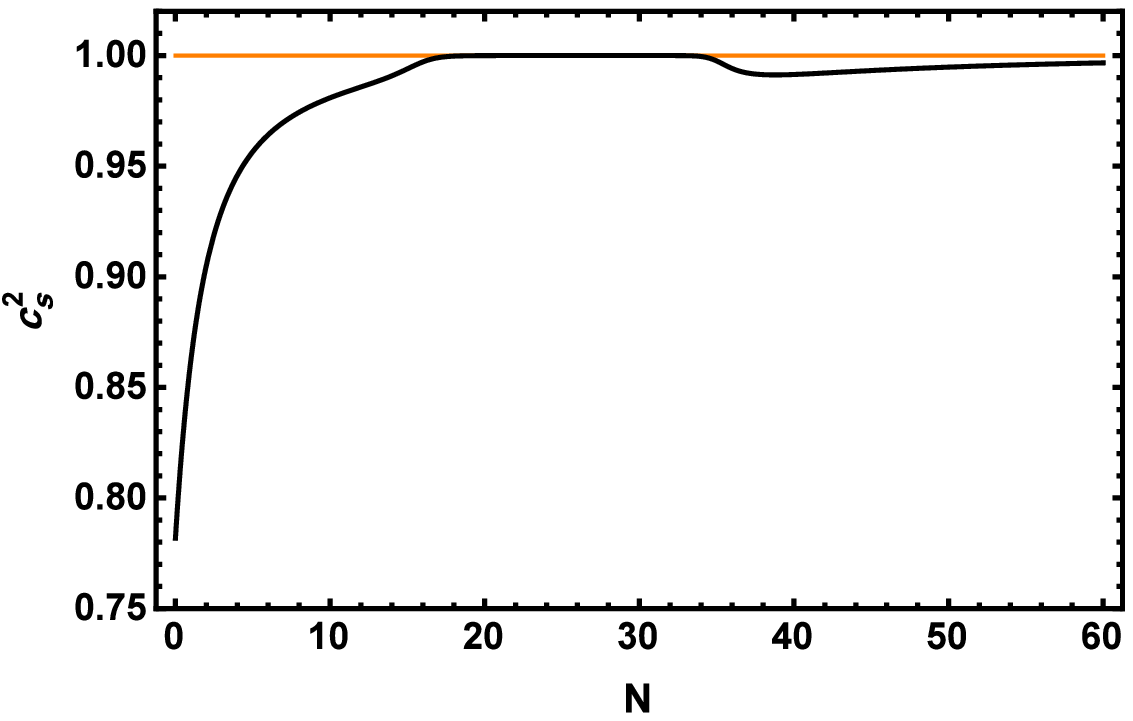}} \hspace{.1cm}
\subfigure{ \includegraphics[width=.48\textwidth]%
{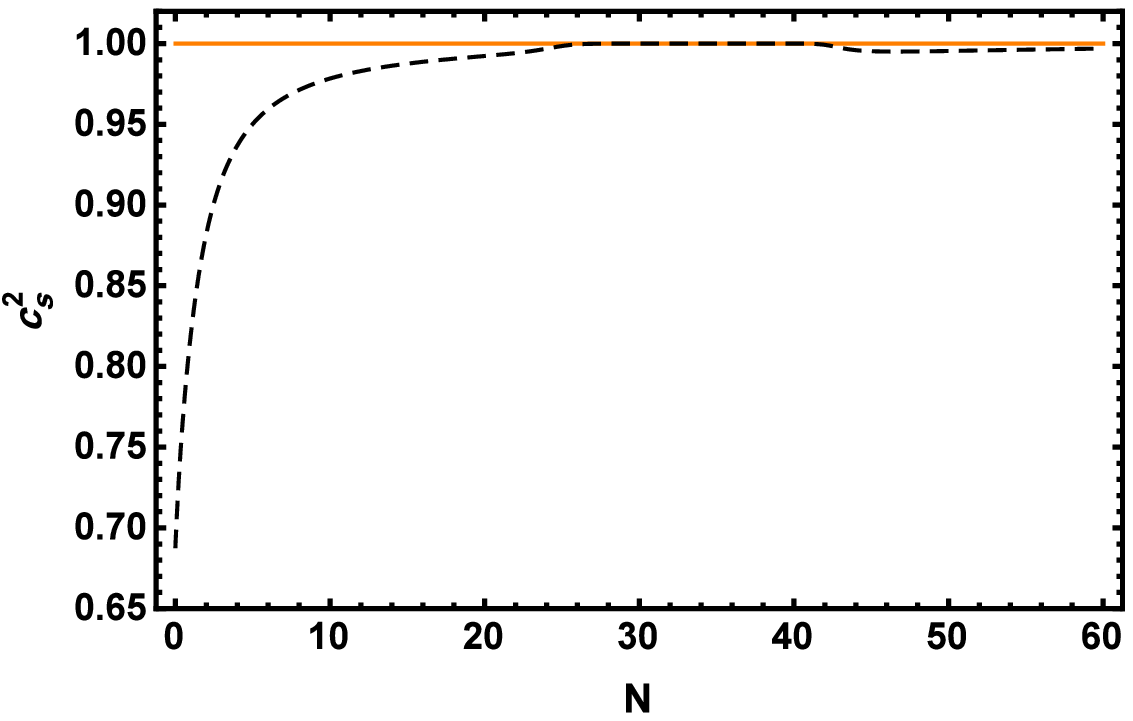}}\hspace{.1cm}
\subfigure{ \includegraphics[width=.48\textwidth]%
{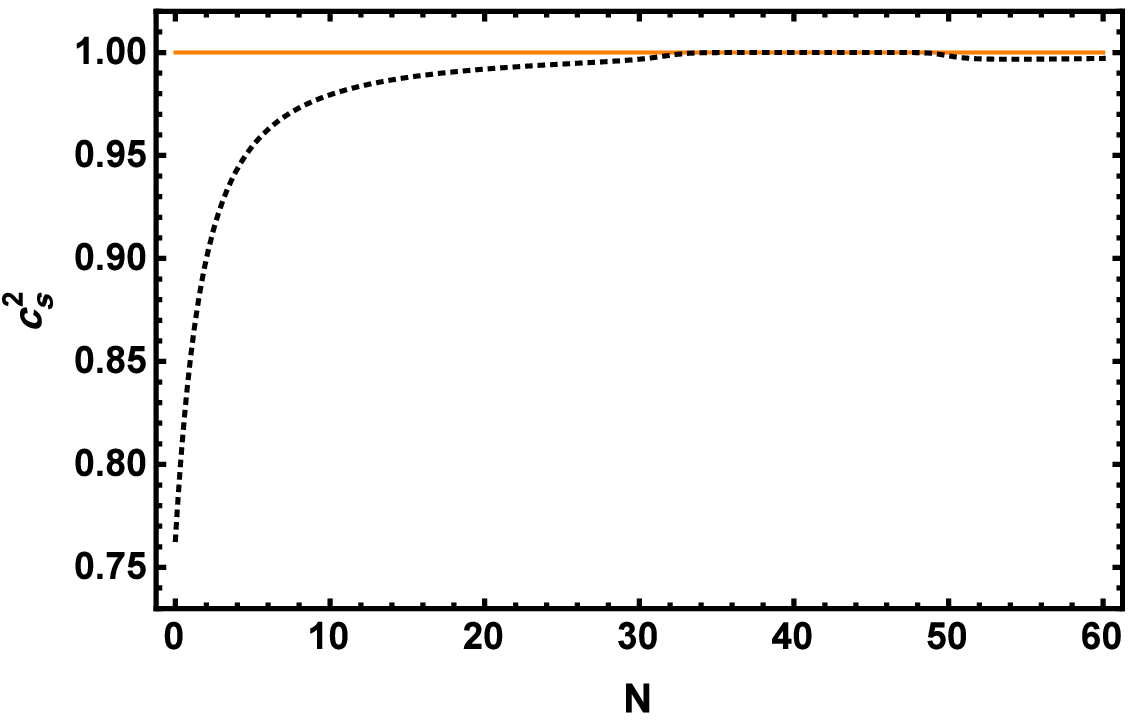}}\hspace{.1cm}
\end{minipage}
\caption{Evolution of the sound speed squared $c_s^2$ versus the $e$-fold number $N$. The solid, dashed, and dotted plots correspond to Cases 1, 2, and 3, respectively.}
\label{fig:cs2}
\end{figure*}

The evolution of the scalar propagation speed squared $c_s^2$ versus $N$ is illustrated in Figure  \ref{fig:cs2}. The figure shows that the sound speed squared always remains subluminal ($c_s^2 < 1$) during the whole period of time domain of interest, and it approaches values very close to unity during the ultra slow-roll phase.

Due to the violation of slow-roll conditions in the plateau-like region, we cannot use Eq. (\ref{PsNatural}) to find the power spectrum during the ultra slow-roll phase, because this equation is obtained by accepting that the slow-roll limit and the condition (\ref{condition}) are valid during inflation. In such a situation, we should compute the exact power spectrum. To this aim, the Mukhanov-Sasaki equation should be solved numerically for all the Fourier modes of interest:

\begin{equation}\label{MS}
 \upsilon^{\prime\prime}_{k}+\left(c_{s}^2 k^2-\frac{z^{\prime\prime}}{z}\right)\upsilon_k=0,
\end{equation}
where a prime denotes a derivative with respect to the conformal time $\eta=\int {a^{-1}dt}$, and

\begin{equation}\label{z}
 \upsilon\equiv z {\cal R}, \hspace{1cm} z=a\sqrt{2Q_s}.
\end{equation}
At the sub-horizon scale $c_{s}k\gg aH$, we can assume that the function $\upsilon$ to be in the Bunch-Davies vacuum, so that its Fourier transform $\upsilon_k$ satisfies \citep{DeFelice:2013ar}

\begin{equation}\label{Bunch}
\upsilon_k\rightarrow\frac{e^{-i c_{s}k\eta}}{\sqrt{2c_s k}}.
\end{equation}
The Mukhanov-Sasaki equation (\ref{MS}) indicates that during inflation, each mode $\upsilon_k$ evolves in time, until this mode stops evolving and reaches a constant value at the super-horizon where $c_{s} k\ll aH$. By solving the Mukhanov-Sasaki equation, numerically, we find the evolution of real and imaginary parts of $\upsilon_k$, then we estimate the scalar power-spectrum of each mode $\upsilon_k$ using the following relation

\begin{equation}\label{PsBunch}
{\cal P}_{s}=\frac{k^3}{2\pi^2}\Big|{\frac{{\upsilon_k}^2}{z^2}}\Big|_{c_{s}k\ll aH}.
\end{equation}
In Figure \ref{fig:Psk}, we plot the actual power spectra obtained from the numerical solution of the Mukhanov-Sasaki equation (\ref{MS}), as a function of the comoving wavenumber $k$. The observational constraints on the primordial curvature power spectrum are also shown. The solid, dashed, and dotted black lines corresponding to the Case1, Case 2, and Case 3, respectively. Figure \ref{fig:Psk} shows that on the large scales when inflation is in the slow-roll phase, the power spectra are in good agreement with the CMB constraints \citep{Akrami:2018odb}. From this figure, we also see that when the ultra slow-roll phase occurs, the power spectra start to increase to the order of ${\cal O}(10^{-2})$  which is large enough to form PBHs.

\begin{figure*}
\begin{minipage}[b]{1\textwidth}
\subfigure{\includegraphics[width=.48\textwidth]%
{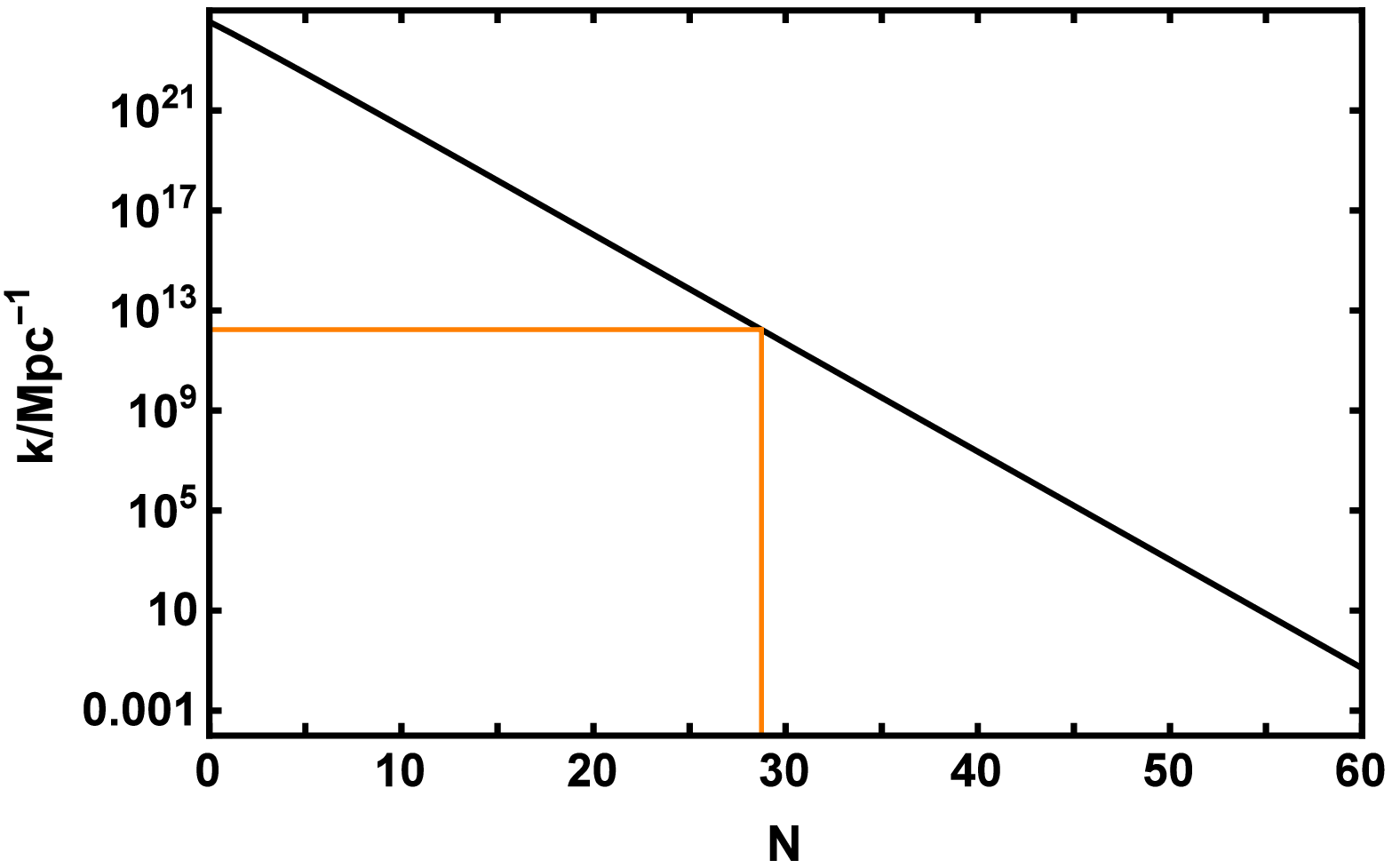}} \hspace{.1cm}
\subfigure{ \includegraphics[width=.48\textwidth]%
{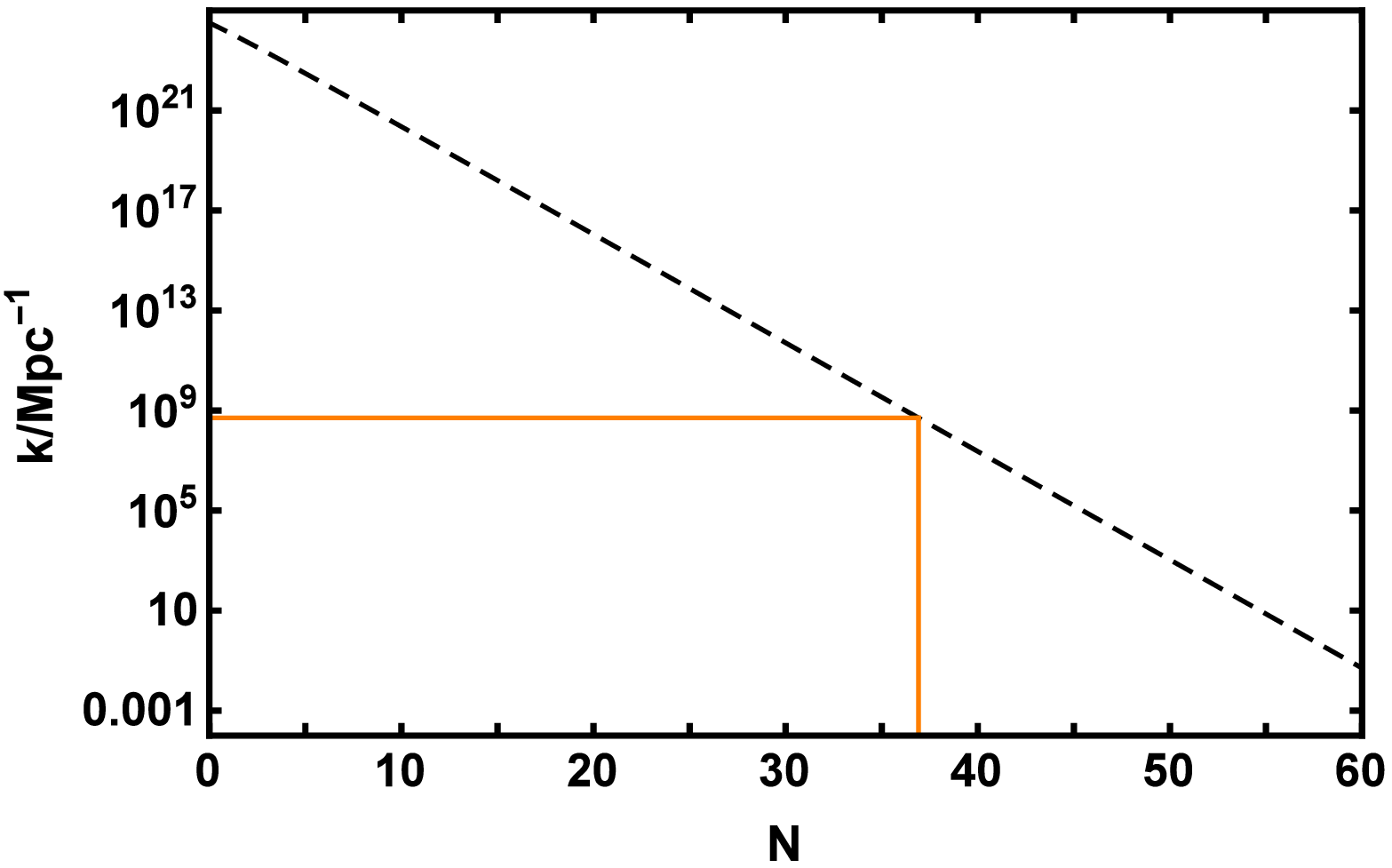}}\hspace{.1cm}
\subfigure{ \includegraphics[width=.48\textwidth]%
{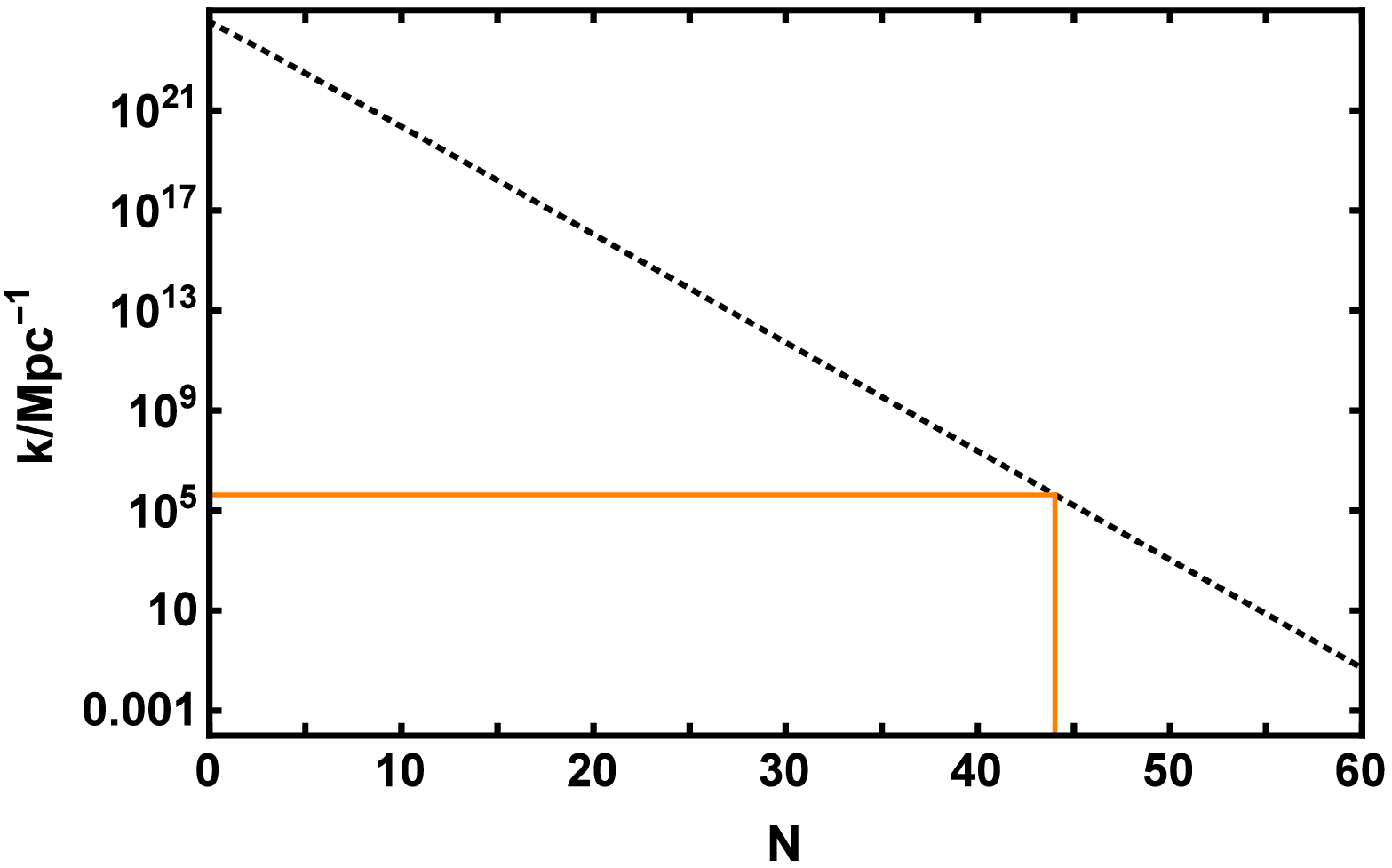}}\hspace{.1cm}
\end{minipage}
\caption{Evolution of the comoving wavenumber $k$ versus the $e$-fold number $N$. The solid, dashed, and dotted plots correspond to Cases 1, 2, and 3, respectively. The orange horizontal and vertical lines specify respectively $k$ and $N$ of the peak of scalar power spectrum for each case.}
\label{fig:kN}
\end{figure*}

From the diagram of the scalar power spectra in Figure  \ref{fig:Psk}, it seems to some extent strange how the comoving wavenumber changes from $k_{\ast}=0.05\,\mathrm{Mpc^{-1}}$ at the horizon crossing with $N_{\ast}=60$ to such rather large values at later epochs at which the peaks of the power spectra appear, while the variations of $e$-folding are not very substantial in this period. In order to explain this exotic behavior, we plot in Figure \ref{fig:kN} the variation of the comoving wavenumber $k$ against of the $e$-fold number $N$ at which that $k$ leaves the Hubble horizon during inflation. We see in this figure that $\log(k)$ has an almost linear relation with $-N$, or, in other words, $k$ is approximately proportional to $\exp(-N)$. Due to this exponential dependency, it is possible for $k$ to change considerably within a few number of $e$-folds. In the figure, we have also specified the comoving wavenumber and its corresponding $e$-fold number at which the peak of the scalar power spectrum takes place for each Case 1, 2, and 3. The numerical values of these quantities are presented in Table \ref{table:NkPs}. In the table, we also report the peak value of the scalar power spectrum. The numbers for $k^{\rm{peak}}$ and ${\cal P}_s^{\rm{peak}}$ in the table are in agreement with those can be seen visually in Figure  \ref{fig:Psk}.

\begin{table*}[ht!]
  \centering
  \caption{The amplitude of the peak of the scalar power spectrum (${\cal P}_s^{\rm{peak}}$) for Cases 1, 2, and 3. $N^{\rm{peak}}$ and $k^{\rm{peak}}$ are the $e$-fold number and comoving wavenumber corresponding to ${\cal P}_s^{\rm{peak}}$, respectively.}
\scalebox{1}[1] {\begin{tabular}{c c c c}
    \hline
    \hline
     \#  &  \,\,\,\,\,\,\,\,\,\,\,\,\,\,\,\,\,\,\,\,\,\,\,\,\,\,\,\,\,\,\,\,\,\,\,\,\,\,\,\,\,\,\,\,\,$N^{\rm{peak}}$ & \,\,\,\,\,\,\,\,\,\,\,\,\,\,\,\,\,\,\,\,\,\,\,\,\,\,\,\,\,\,\,\,\,\,\, $k^{\rm{peak}}/\mathrm{Mpc^{-1}}$  &  \,\,\,\,\,\,\,\,\,\,\,\,\,\,\,\,\,\,\,\,\,\,\,\,\,\,\,\,\,\,\,\,\,\,\,\,\,\,\, ${\cal P}_s^{\rm{peak}}$\\
    \hline
    Case 1 &  \,\,\,\,\,\,\,\,\,\,\,\,\,\,\,\,\,\,\,\,\,\,\,\,\,\,\,\,\,\,\,\,\,\,\, $28.74$  & \,\,\,\,\,\,\,\,\,\,\,\,\,\,\,\,\,\,\,\,\,\,\,\,\,\,\,\,\,\,\,\,\,\,\,  $1.714\times10^{12}$   & \,\,\,\,\,\,\,\,\,\,\,\,\,\,\,\,\,\,\,\,\,\,\,\,\,\,\,\,\,\,\,\,\,\,\, $0.0316$ \\
    Case 2 & \,\,\,\,\,\,\,\,\,\,\,\,\,\,\,\,\,\,\,\,\,\,\,\,\,\,\,\,\,\,\,\,\,\,\,$36.91$  &  \,\,\,\,\,\,\,\,\,\,\,\,\,\,\,\,\,\,\,\,\,\,\,\,\,\,\,\,\,\,\,\,\,\,\, $5.071\times10^{8}$    & \,\,\,\,\,\,\,\,\,\,\,\,\,\,\,\,\,\,\,\,\,\,\,\,\,\,\,\,\,\,\,\,\,\,\,$0.0388$\\
    Case 3 & \,\,\,\,\,\,\,\,\,\,\,\,\,\,\,\,\,\,\,\,\,\,\,\,\,\,\,\,\,\,\,\,\,\,\,$44.01$  & \,\,\,\,\,\,\,\,\,\,\,\,\,\,\,\,\,\,\,\,\,\,\,\,\,\,\,\,\,\,\,\,\,\,\, $4.259\times10^{5}$    & \,\,\,\,\,\,\,\,\,\,\,\,\,\,\,\,\,\,\,\,\,\,\,\,\,\,\,\,\,\,\,\,\,\,\,$0.0460$ \\
    \hline
    \end{tabular}}
  \label{table:NkPs}
\end{table*}

\begin{figure*}[t]
\begin{center}
\scalebox{0.9}[0.9]{\includegraphics{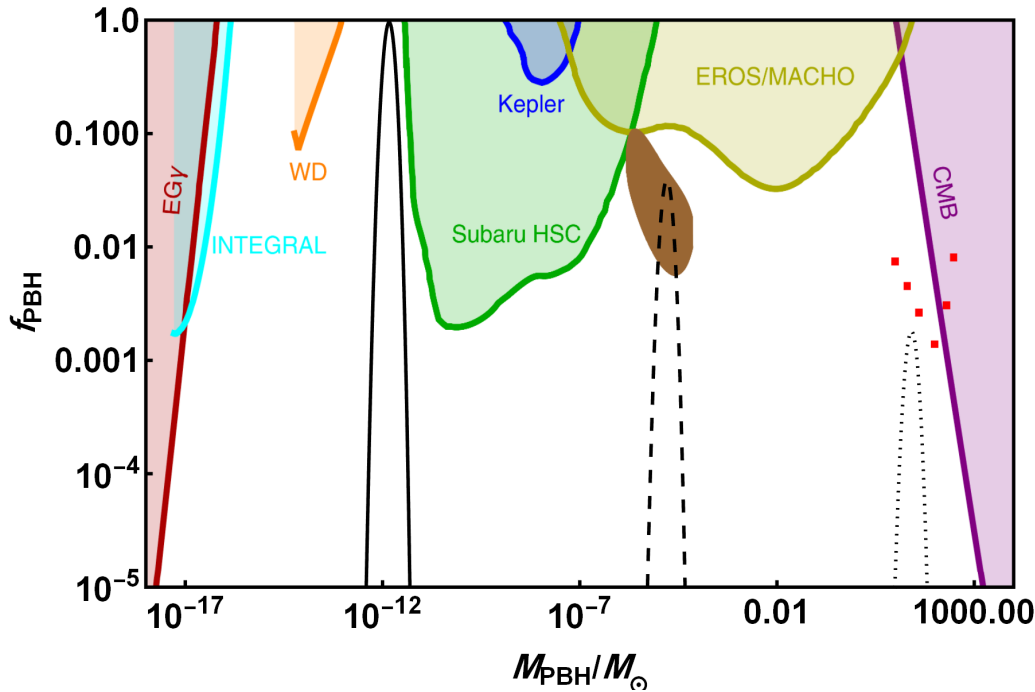}}
\caption{The fractional abundance of PBHs as a function of PBH mass for Case 1 (solid line), Case 2 (dashed line), and Case 3 (dotted line). The red points show the potential upper bounds on the PBH abundance from requiring that the merger rate of PBHs does not exceed the upper limit on the LIGO merger rate \citep{Ali-Haimoud:2017rtz}. The brown-shaded region represents the allowed PBH abundance from the ultrashort-timescale microlensing events in the OGLE data \citep{mroz2017no, Niikura:2017zjd}. The other shaded regions indicate the current observational constraints on the abundance of PBHs: extragalactic gamma-rays from PBH evaporation (EG$\gamma$) \citep{Carr:2009jm}, galactic center 511 keV $\gamma$-ray line (INTEGRAL) \citep{Laha:2019ssq}, white dwarfs explosion (WD) \citep{Graham:2015apa}, microlensing events with Subaru HSC (Subaru HSC) \citep{Niikura:2017zjd}, with the Kepler satellite (Kepler) \citep{Griest:2013esa}, with EROS/MACHO (EROS/MACHO) \citep{Tisserand:2006zx}, and accretion constraints from CMB (CMB) \citep{Ali-Haimoud:2016mbv}.
}
\label{fig:fPBH}
\end{center}
\end{figure*}

Substituting the actual power spectrum obtained by solving the Mukhanov-Sasaki equation, into Eq. (\ref{sigma2}) and then calculating the integral, then using Eqs. (\ref{mass}), (\ref{beta}), and (\ref{fPBH}), we can find the primordial black holes abundance for the parameter values given in Table \ref{Table1}. The results are shown in Figure  \ref{fig:fPBH} and Table \ref{Table2}. Our analysis indicates that in Case 1, our model generates PBHs with mass $M\simeq 1.47\times10^{-12} M_\odot$ and $f_{\rm{PBH}}\simeq 0.9563$ which means that this class of PBHs constitute around $0.96\%$ of DM content of the universe. Therefore, the formed PBHs in Case 1, can be taken as a suitable candidate for DM.

For the parameters set of Case 2, the model predicts PBHs with the mass around $ {\cal O}(10^{-5})M_\odot$ which the peak of $f_{\rm PBH}$ places on the inferred region of the PBH abundance by the ultrashort-timescale microlensing events in OGLE data. Therefore, one can consider these PBHs as a source of these microlensing events.

In Case 3, the produced PBHs mass is $M\simeq 25.70 M_\odot$ with $f_{\rm {PBH}}\simeq0.0018$. Figure  \ref{fig:fPBH} indicates this class of PBHs can satisfy the constraint from the upper limit on the LIGO merger rate.

\section{Secondary gravitational waves}
\label{sec:sgws}

\begin{figure*}[t]
\begin{center}
\scalebox{0.45}[0.45]{\includegraphics{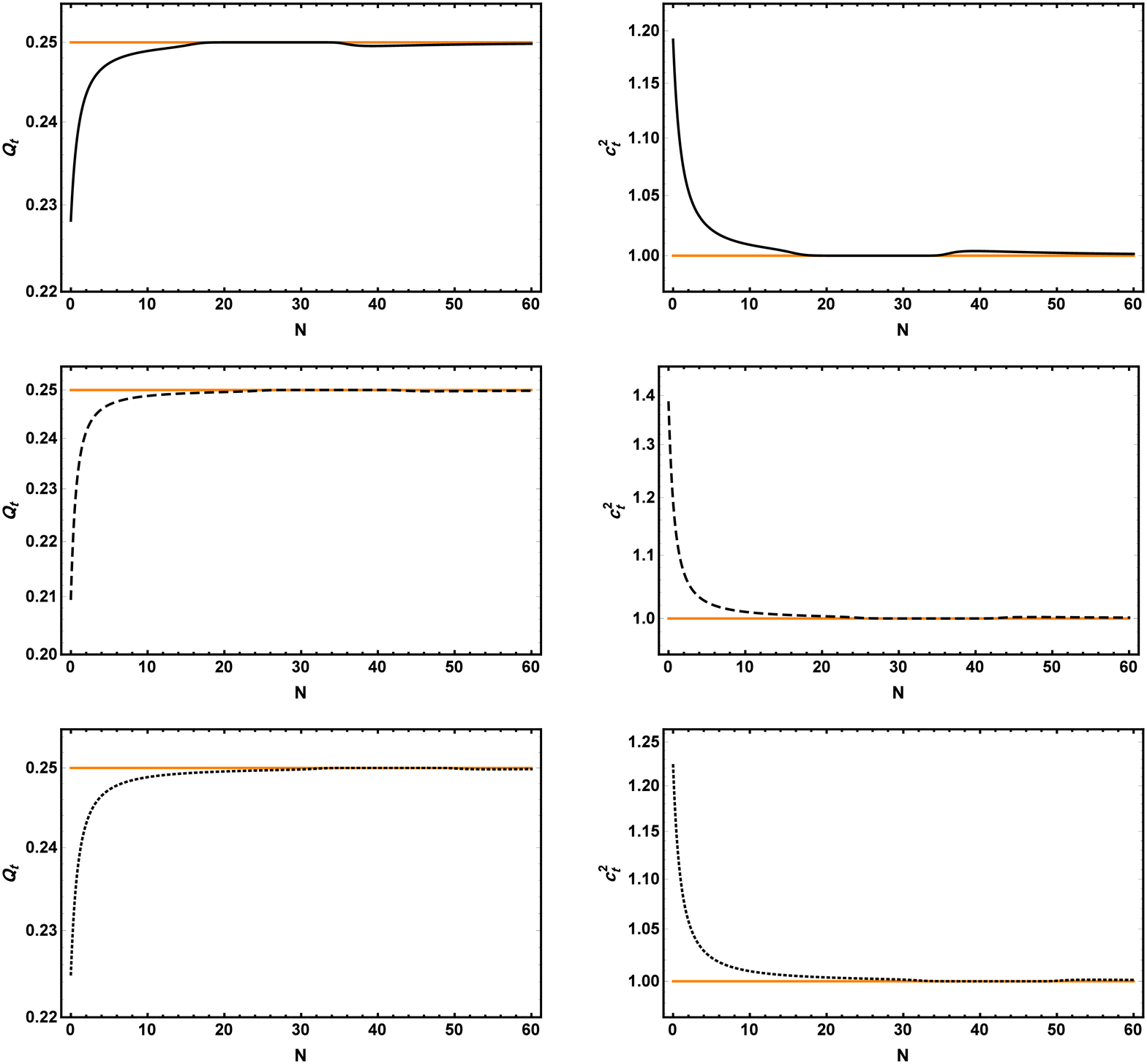}}
\caption{Evolution of $Qt$ (left) and $c_t^2$ (right) versus $N$. The solid, dashed, and dotted plots correspond to Cases 1, 2, and 3, respectively.}
\label{fig:Qtct2}
\end{center}
\end{figure*}

Production of the primordial perturbation from inflation may induce the generation of secondary GWs. These GWs are a significant criteria to check the validity of inflationary models, because they can be tested through the data of some designed GWs detectors. In this section, we aim to calculate the secondary GWs in our nonminimal derivative inflationary setup. For this purpose, we first decompose the transverse and traceless tensor perturbation $h_{ij}$ into two independent polarization modes, as $h_{ij}=h_{+}e_{ij}^{+}+h_{\times}e_{ij}^{\times}$. The tensors modes $e_{ij}^{\lambda}$ with $\lambda=+,\times$, satisfy the orthogonality relations $e_{ij}^{+}(\boldsymbol{k})e_{ij}^{+}(-\boldsymbol{k})^{*}=2$, $e_{ij}^{\times}(\boldsymbol{k})e_{ij}^{\times}(-\boldsymbol{k})^{*}=2$, and $e_{ij}^{+}(\boldsymbol{k})e_{ij}^{\times}(-\boldsymbol{k})^{*}=0$ in Fourier space. The second-order action for the tensor perturbation in our model takes the following form \citep{DeFelice:2011uc,Gao:2011qe,Kobayashi:2011nu,Tsujikawa:2013ila}

\begin{equation}
 \label{St2}
 S_{t}^{(2)}={\sum\limits_{\lambda=+,\times}^{} {\int dtd^{3}xa^{3}Q_{t}\left[\dot{h}_{\lambda}^{2}-\frac{c_{t}^{2}}{a^{2}}\left(\partial h_{\lambda}\right)^{2}\right]}},
\end{equation}
where $Q_t$ and $c_t^2$ are given by Eqs. \eqref{Qt} and \eqref{ct2}, respectively. In Figure  \ref{fig:Qtct2}, we plot $Q_t$ (left panel) and $c_t^2$ (right panel), versus number of $e$-folding during inflation in our model.
Figure  \ref{fig:Qtct2}, shows that the tensor propagation speed squared $c_{t}^2$ is greater than unity during some periods of the time domain of interest.
In general, the theories with noncanonical kinetic terms typically show a superluminal behavior for the tensor modes \citep{Izumi:2014loa,Reall:2014pwa,deser2013acausality,Deser:2014hga,Deser:2013qza,Minamitsuji:2015nca}. The nonminimal derivative coupling model is one such theory. This superluminal behavior necessarily does not threaten causality, because the Lorentz symmetry on the FRW background is explicitly broken \citep{Tsujikawa:2012mk}. In these theories, a different notion of causality than the standard local Lorentz invariant theories is needed \citep{burrage2012chronology}. \citet{Minamitsuji:2015nca} has studied the causal structure in nonminimal derivative coupling scenario and argued that the superluminality does not necessarily mean the violation of causality, and declared that this subject requires the further careful studies. If a superluminal mode exists, the appearance of the closed time-like curves (CTCs) is a problem that can occurs \citep{Minamitsuji:2015nca,burrage2012chronology, ohashi2012potential,Izumi:2014loa,Burrage:2011cr}. In \citet{Burrage:2011cr}, it has been proposed that the construction of CTCs is not possible in the models with the noncanonical kinetic terms, since at the onset of the formation of a CTC, the scalar field becomes strongly coupled. Besides, it is worth mentioning that, as argued in \citet{Ezquiaga:2017ekz,Kase:2018aps}, the nonminimal derivative coupling theory cannot serve as a dark energy model, after the observation of GW170817 detected by the LIGO-VIRGO Collaboration on August 17, 2017 \citep{TheLIGOScientific:2017qsa}, which places a bound on the GW speed at low redshifts. Therefore, we eliminate any late-time application of this theory, and confine ourselves to its implications for the early universe.

\begin{figure*}[t]
\begin{center}
\scalebox{0.9}[0.9]{\includegraphics{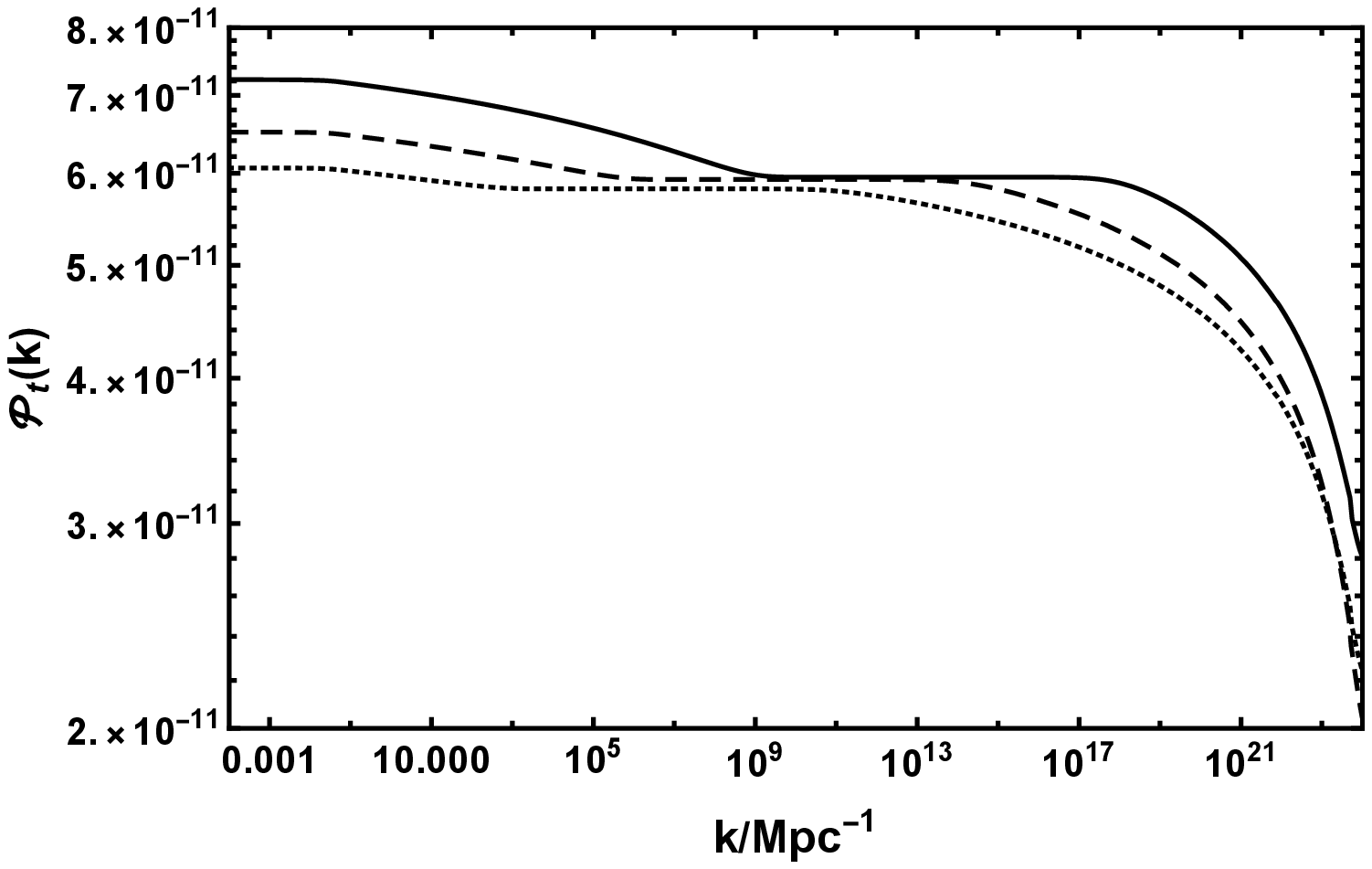}}
\caption{The tensor power spectra found by solving the Mukhanov-Sasaki equation, numerically, as a function of the comoving wavenumber $k$ for Case 1 (solid line), Case 2 (dashed line), and Case 3 (dotted line).}
\label{fig:Ptk}
\end{center}
\end{figure*}

Applying the same procedure that was followed for the scalar perturbations, we can compute the exact tensor power spectrum  during inflation. After deriving the evolution of real and imaginary parts of $u_k$ for each mode $k$, we estimate the tensor power-spectrum using the expression ${\cal P}_{t}=4k^3|{h_{\lambda}}|^2/2\pi^2$ in which $u_k\equiv z h_{\lambda}$ and $z_t=a\sqrt{2Q_t}$ \citep{de2011primordial}. The actual tensor power spectra as a function of the comoving wavenumber $k$ is plotted in Figure  \ref{fig:Ptk}. Our numerical calculations imply that in our model, the amplitude of tensor power spectrum is $\mathcal{P}_{t}\sim 2.0-7.2\times10^{-11}$ during the time domain of interest, and it decreases as we go towards the smaller scales (larger values of $k$). In the ultra slow-roll phase, the tensor power spectrum becomes very flat and to explain this behavior, we note that the tensor power spectrum depends on the energy scale of the universe, and it in turn is related to the Hubble parameter through the first Freidmann equation \eqref{FR1:eq}. In addition, the Hubble parameter varies very slowly with time in the ultra slow-roll regime, due to the friction coming from the peak of the nonminimal derivative coupling term. Therefore, a plateau-like region appears in the tensor power spectrum for the values of $k$ corresponding to the ultra slow-roll regime.

In Figure  \ref{fig:Qtct2}, we see that the quantities of $Q_{t}$ and  $c_{t}^{2}$ for a considerable period of the time domain of interest are approximately as $Q_{t}\approx1/4$ and $c_{t}^{2}\approx1$ in our model, and so the action \eqref{St2} simply reduces to

\begin{equation}
 \label{St2Einstein}
 S_{t}^{(2)} \approx \frac{1}{4}{\sum\limits_{\lambda=+,\times}^{}{\int dtd^{3}xa^{3}\left[\dot{h}_{\lambda}^{2}-\frac{1}{a^{2}}\left(\partial h_{\lambda}\right)^{2}\right]}},
\end{equation}
which is the second-order action for the tensor perturbations in the Einstein gravity \citep{Boubekeur:2008kn}.

Therefore, we conclude that up to a good approximation, the dynamics of GWs in our model is similar to their dynamics in the Einstein theory. We besides suppose that this approximation is also preserved throughout the subsequent cosmological evolutions, and so we are allowed to apply the formulation based on the Einstein general relativity to evaluate the dynamics of GWs in our model. It has been shown that in this formulation, the fractional energy density of the induced GWs in the radiation dominated era is given by \citep{Lu:2019sti}

\begin{align}
\label{OmegaGW}
& \Omega_{GW}(k,\eta)=\frac{1}{6}\left(\frac{k}{aH}\right)^{2}\int_{0}^{\infty}dv
\int_{|1-v|}^{|1+v|}
\nonumber
\\
& du\left(\frac{4v^{2}-\left(1-u^{2}+v^{2}\right)^{2}}{4uv}\right)^{2}\overline{I_{RD}^{2}(u,v,x)}\mathcal{P}_{s}(ku)\mathcal{P}_{s}(kv),
\end{align}
where $\eta$ is the conformal time, and the time average of the source terms is

\begin{align}
 &\overline{I_{RD}^{2}(u,v,x\to\infty)}= \nonumber \\
 & \frac{1}{2x^{2}}\Bigg[\Bigg(\frac{3\pi\left(u^{2}+v^{2}-3\right)^{2}\Theta\left(u+v-\sqrt{3}\right)}{4u^{3}v^{3}}
 \nonumber
 \\
 & + \frac{T_{c}(u,v,1)}{9}\Bigg)^{2}
  +\left(\frac{\tilde{T}_{s}(u,v,1)}{9}\right)^{2}\Bigg],
 \label{IRD2b}
\end{align}
where the expressions of the functions $T_c$ and $\tilde{T}_{s}$ are given in Appendix \ref{appendix:functions}.

Due to the fact that GWs behave like radiation, the present energy densities of GWs are related to their values well after the horizon reentry in the radiation dominated era,

\begin{equation}
 \label{OmegaGW0}
 \Omega_{GW}\left(k,\eta_{0}\right)=\Omega_{GW}(k,\eta)\frac{\Omega_{r0}}{\Omega_{r}(\eta)},
\end{equation}
where $\Omega_r$ is the fractional energy density of radiation, and the subscript $0$ refers to the present epoch. In this work, we have taken the current radiation density parameter as $\Omega_{r0}h^{2}\simeq4.2\times10^{-5}$ \citep{Cai:2019bmk,Fu:2019vqc,Fu:2020lob}. It should be noted that in Eq. \eqref{OmegaGW0}, the conformal time $\eta\gg\eta_{k}$ is taken to be earlier than the epoch of matter radiation equality, and of course late enough so that $\Omega_{GW}(k,\eta)$ converges to a constant value.

\begin{figure*}
\begin{center}
\scalebox{0.9}[0.9]{\includegraphics{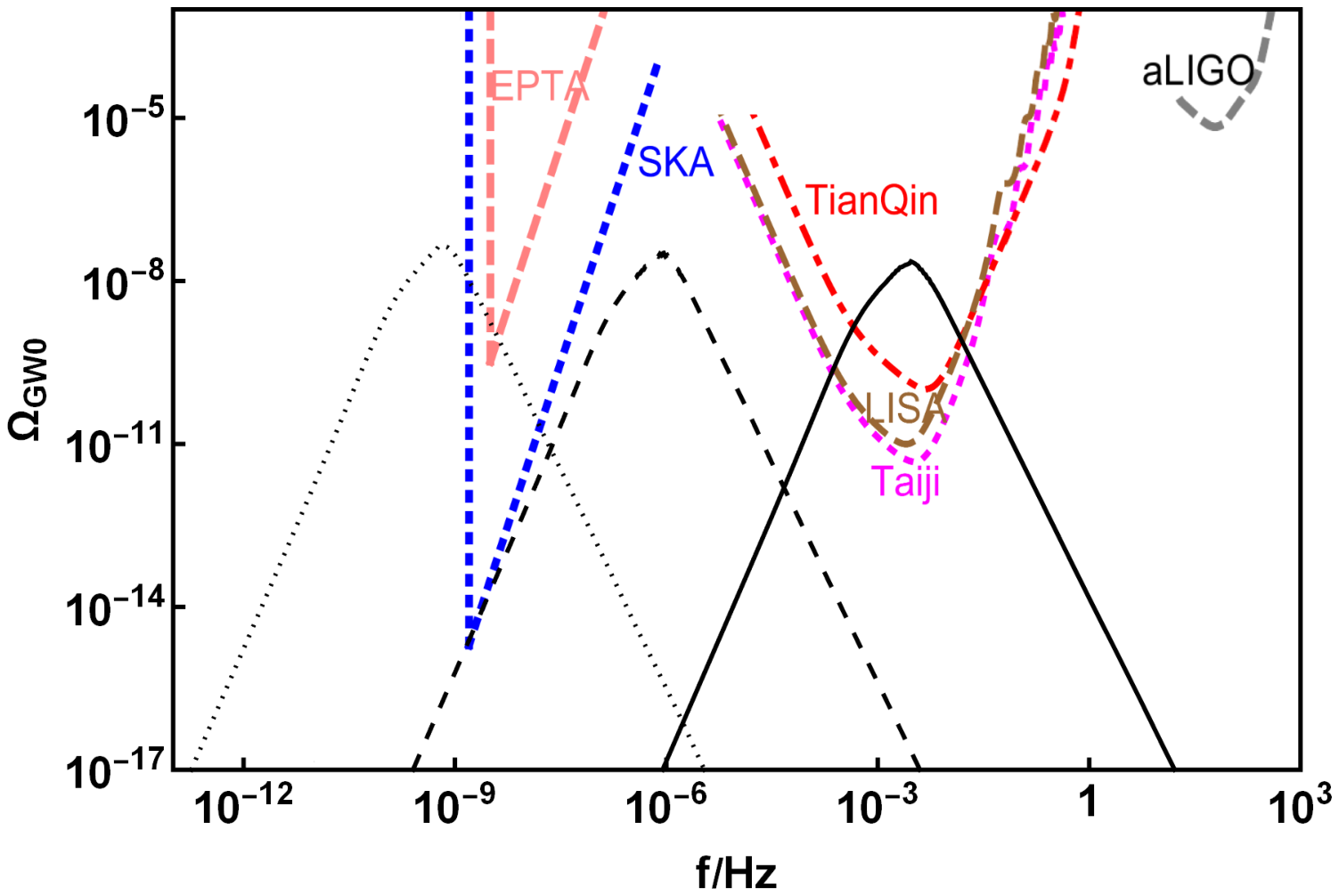}}
\caption{The present fractional energy density of the secondary GWs in terms of frequency. The solid, dashed, and dotted plots correspond to Cases 1, 2, and 3, respectively.}
\label{fig:OmegaGW0}
\end{center}
\end{figure*}

We use Eq. \eqref{OmegaGW0} and plot the present fractional energy density of the secondary GWs for the three cases of our model as presented in Figure  \ref{fig:OmegaGW0}.
To compare the predictions of our model with observations, the sensitivity curves of European PTA (EPTA) \citep{Ferdman:2010xq,Hobbs:2009yy,McLaughlin:2013ira}, the Square Kilometer Array (SKA) \citep{Moore:2014lga}, Advanced Laser Interferometer Gravitational Wave Observatory (aLIGO) \citep{Harry:2010zz,TheLIGOScientific:2014jea}, Laser Interferometer Space Antenna (LISA) \citep{Danzmann:1997hm,Audley:2017drz}, TaiJi \citep{Hu:2017mde}, and TianQin \citep{Luo:2015ght}, are also specified in the figure. As we see in the figure, the prediction of Cases 1, 2, and 3 of our model have peaks for $\Omega_{GW0}$ at different frequencies, while the height of the peaks is almost similar for all of them and are of order $10^{-8}$. For Case 1, the peak appears at the frequency $2.953\times10^{-3}Hz$, and it lies inside the sensitivity regions of LISA, TaiJi, and TianQin. In Case 2, the peak takes place at $8.017\times10^{-7}Hz$, but the prediction of this case lies outside the sensitivity curves of the mentioned GWs detectors. The peak of Case 3 occurs at $5.848\times10^{-10}Hz$, and it is located within the sensitivity region of EPTA and SKA. Since the predictions of our model in some cases can lie inside the sensitivity of the GWs detectors, therefore the viability of our model can be checked when the data of these detectors are released.

\section{Conclusions}\label{sec:con}

In this work, we investigated PBHs formation scenario in a single field model of inflation with a nonminimal field derivative coupling to gravity described by the action (\ref{action}), which belongs to the family of Horndeski theory. The nonminimal derivative coupling to gravity provides a gravitationally enhanced friction mechanism that works for some special potentials. Therefore, we considered the potential of the natural inflation $V(\phi)=\Lambda^4\big[1+\cos(\phi/f)\big]$, where $\Lambda$ and $f$ are constants with dimensions of mass.
Note that the predictions of the natural potential in the standard scenario are compatible with only the 95\% CL constraints of Planck 2018 TT+lowE data \citep{Akrami:2018odb}. This motivated us to investigate the natural inflation in the nonminimal derivative coupling scenario to see whether its predictions can be improved in light of the current CMB data.
With the coupling parameter $g(\phi)$ in the form of Eqs. (\ref{g})-(\ref{gII}), for $\alpha=0$, $f=0.2$, and $\lambda\equiv f^{1+\alpha} \Lambda^4/M^{1+\alpha}=20$ we fixed $\Lambda$, using the normalization of the power spectrum at the epoch of horizon crossing with the $e$-fold number $N_{*}=60$ as ${\cal P}_{s*}\equiv{\cal P}_s \big|_{N=N_*}\simeq 2.1\times10^{-9}$, and then by fine-tuning of the model parameters, we found three parameter sets (see Table \ref{Table1}) which produce an ultra slow-roll phase in which the evolution of inflaton slows down enough to form PBHs by increasing friction. By solving the dynamical equations numerically, we plot the evolution of the scalar field $\phi$, the first slow-roll parameter $\varepsilon$, the second slow-roll parameter $\delta_{\phi}$, the expression ${g_{,\phi}H\dot{\phi}}/{\cal A}$ and the scalar propagation speed squared $c_s^2$, as functions of $e$-fold number $N$ in Figs. \ref{fig:phi}-\ref{fig:cs2}. From Figure  \ref{fig:SRparameters} it is clear that in the ultra slow-roll phase, the first slow-roll condition is preserved ($\varepsilon\ll1$), but the second one is violated ($\left|\delta_{\phi}\right|\gtrsim1$). Consequently, to compute the power spectra of curvature perturbations, we solved the Mukhanov-Sasaki equation for three cases summarized in Table \ref{Table1}, numerically. We showed that the resulting power spectra have large enough peaks to produce PBHs on the small scales and satisfy the Planck observational constraints on the large scales (see Figure  \ref{fig:Psk} and Table \ref{table:NkPs}).
Moreover, we estimated the inflationary observables in our model and showed that the values of $r$, and $dn_s/d\ln k$ predicted by Cases 1 and 2, satisfy  the $68\%$ CL constraints of Planck 2018 TT+lowE data \citep{Akrami:2018odb}, while the values of $n_s$ are in agreement with the 95\% CL constraints of these observational data. For Case 3, the values  of $n_s$ and $r$ fulfills the $68\%$ CL constraints of Planck 2018 TT+lowE data \citep{Akrami:2018odb}, while the value of $dn_s/d\ln k$ is consistent with the 95\% CL constraint of these observational data (see Table \ref{Table2}). Thus, we conclude that in the framework of the nonminimal derivative coupling, we can improve the consistency of the inflationary observables of the natural potential in light of the Planck 2018 results \citep{Akrami:2018odb} in comparison with its results in the setup of standard inflationary scenario.

Applying the Press-Schechter formalism and using the exact power spectra, we obtained PBHs with masses ${\cal O}(10^{-12})M_\odot$ (Case 1), ${\cal O}(10^{-5})M_\odot$ (Case 2), and ${\cal O}(10)M_\odot$ (Case 3), which can explain around $96\%$ of DM, the ultrashort-timescale microlensing events in OGLE data and the LIGO events, respectively (see Figure  \ref{fig:fPBH} and Table \ref{Table2}). We realized that the produced PBHs in Case 1 can be taken as a good candidate for DM.

We plotted the exact tensor power spectra as a function of the comoving wavenumber $k$, in Figure \ref{fig:Ptk}. Our findings imply that in our scenario, the amplitude of tensor power spectrum is $\mathcal{P}_{t}\sim 2.0-7.2\times10^{-11}$ during the time domain of interest, and it decreases as we go towards the smaller scales (larger values of $k$). We further showed that the tensor power spectrum of our model exhibits a plateau-like region in the comoving wavenumbers related to the ultra slow-roll phase.

We moreover studied the secondary GWs induced due to the PBHs formation in our setup. In particular, we estimated the present fractional energy density ($\Omega_{GW0}$) for the three parameter sets of our model, and showed that these cases result in peaks for $\Omega_{GW0}$ in different frequencies. The peak height is of order $10^{-8}$ and it is almost identical for all the cases. The peaks for the Cases 1, 2, and 3 take place at the frequencies $2.953\times10^{-3}Hz$, $8.017\times10^{-7}Hz$, and $5.848\times10^{-10}Hz$, respectively. The peak of $\Omega_{GW0}$ for Case 1 lies in the sensitivity region of LISA, TaiJi, and TianQin, and for Case 3 in the sensitivity regions of EPTA and SKA, while for Case 2, the peak cannot be placed inside the sensitivity regions. Consequently, our model is capable to be tested in light of GWs observations when the data of these detectors are released.

\acknowledgments

The authors thank the referee for his/her valuable comments.

\appendix

\section{Analytical functions}
\label{appendix:functions}

In this appendix, we present the expressions of the analytical functions that are used in Eq. \eqref{IRD2b} for time average of the source terms.

The functions $T_{c}$ and $T_{s}$ are defined in the following forms \citep{Lu:2019sti}:

\begin{align}
T_{c}= & -\frac{27}{8u^{3}v^{3}x^{4}}\Bigg\{-48uvx^{2}\cos\left(\frac{ux}{\sqrt{3}}\right)\cos\left(\frac{vx}{\sqrt{3}}\right)\left(3\sin(x)+x\cos(x)\right)+
\nonumber
\\
& 48\sqrt{3}x^{2}\cos(x)\left(v\sin\left(\frac{ux}{\sqrt{3}}\right)\cos\left(\frac{vx}{\sqrt{3}}\right)+u\cos\left(\frac{ux}{\sqrt{3}}\right)\sin\left(\frac{vx}{\sqrt{3}}\right)\right)+
\nonumber
\\
& 8\sqrt{3}x\sin(x)\Bigg[v\left(18-x^{2}\left(u^{2}-v^{2}+3\right)\right)\sin\left(\frac{ux}{\sqrt{3}}\right)\cos\left(\frac{vx}{\sqrt{3}}\right)+
\nonumber
\\
& u\left(18-x^{2}\left(-u^{2}+v^{2}+3\right)\right)\cos\left(\frac{ux}{\sqrt{3}}\right)\sin\left(\frac{vx}{\sqrt{3}}\right)\Bigg]+
\nonumber
\\
& 24x\cos(x)\left(x^{2}\left(-u^{2}-v^{2}+3\right)-6\right)\sin\left(\frac{ux}{\sqrt{3}}\right)\sin\left(\frac{vx}{\sqrt{3}}\right)+
\nonumber
\\
& 24\sin(x)\left(x^{2}\left(u^{2}+v^{2}+3\right)-18\right)\sin\left(\frac{ux}{\sqrt{3}}\right)\sin\left(\frac{vx}{\sqrt{3}}\right)\Bigg\}
\nonumber
\\
& -\frac{\left(27\left(u^{2}+v^{2}-3\right)^{2}\right)}{4u^{3}v^{3}}\Bigg\{\text{Si}\left[\left(\frac{u-v}{\sqrt{3}}+1\right)x\right]-\text{Si}\left[\left(\frac{u+v}{\sqrt{3}}+1\right)x\right]
\nonumber
\\
& +\text{Si}\left[\left(1-\frac{u-v}{\sqrt{3}}\right)x\right]-\text{Si}\left[\left(1-\frac{u+v}{\sqrt{3}}\right)x\right]\Bigg\}
\label{Tc}
\end{align}

\begin{align}
T_{s}= & \frac{27}{8u^{3}v^{3}x^{4}}\Bigg\{48uvx^{2}\cos\left(\frac{ux}{\sqrt{3}}\right)\cos\left(\frac{vx}{\sqrt{3}}\right)\left(x\sin(x)-3\cos(x)\right)-
\nonumber
\\
& 48\sqrt{3}x^{2}\sin(x)\left(v\sin\left(\frac{ux}{\sqrt{3}}\right)\cos\left(\frac{vx}{\sqrt{3}}\right)+u\cos\left(\frac{ux}{\sqrt{3}}\right)\sin\left(\frac{vx}{\sqrt{3}}\right)\right)+
\nonumber
\\
& 8\sqrt{3}x\cos(x)\Bigg[v\left(18-x^{2}\left(u^{2}-v^{2}+3\right)\right)\sin\left(\frac{ux}{\sqrt{3}}\right)\cos\left(\frac{vx}{\sqrt{3}}\right)+
\nonumber
\\
& u\left(18-x^{2}\left(-u^{2}+v^{2}+3\right)\right)\cos\left(\frac{ux}{\sqrt{3}}\right)\sin\left(\frac{vx}{\sqrt{3}}\right)\Bigg]+
\nonumber
\\
& 24x\sin(x)\left(6-x^{2}\left(-u^{2}-v^{2}+3\right)\right)\sin\left(\frac{ux}{\sqrt{3}}\right)\sin\left(\frac{vx}{\sqrt{3}}\right)+
\nonumber
\\
& 24\cos(x)\left(x^{2}\left(u^{2}+v^{2}+3\right)-18\right)\sin\left(\frac{ux}{\sqrt{3}}\right)\sin\left(\frac{vx}{\sqrt{3}}\right)\Bigg\}-\frac{27\left(u^{2}+v^{2}-3\right)}{u^{2}v^{2}}+
\nonumber
\\
& \frac{\left(27\left(u^{2}+v^{2}-3\right)^{2}\right)}{4u^{3}v^{3}}\Bigg\{-\text{Ci}\left[\left|1-\frac{u+v}{\sqrt{3}}\right|x\right]+\ln\left|\frac{3-(u+v)^{2}}{3-(u-v)^{2}}\right|+
\nonumber
\\
& \text{Ci}\left[\left(\frac{u-v}{\sqrt{3}}+1\right)x\right]-\text{Ci}\left[\left(\frac{u+v}{\sqrt{3}}+1\right)x\right]+\text{Ci}\left[\left(1-\frac{u-v}{\sqrt{3}}\right)x\right]\Bigg\}
\label{Ts}
\end{align}
In these equations, we have used the sine-integral $\text{Si}(x)$ and cosine-integral $\text{Ci}(x)$ functions which are defined as

\begin{equation}
 \label{SiCi}
 \text{Si}(x)=\int_{0}^{x}\frac{\sin(y)}{y}dy,\qquad\text{Ci}(x)=-\int_{x}^{\infty}\frac{\cos(y)}{y}dy.
\end{equation}

In addition, in Eq. \eqref{IRD2b}, we have used the function $\tilde{T}_{s}(u,v,1)$ which is expressed as

\begin{equation}
 \label{Tst}
 \tilde{T}_{s}(u,v,1)=T_{s}(u,v,1)+\frac{27\left(u^{2}+v^{2}-3\right)}{u^{2}v^{2}}
 -\frac{27\left(u^{2}+v^{2}-3\right)^{2}}{4u^{3}v^{3}}\ln\left|\frac{3-(u+v)^{2}}{3-(u-v)^{2}}\right|.
\end{equation}


% \bibliography{bibliography}{}
%\bibliographystyle{aasjournal}



\end{document}